\documentclass[12pt,a4paper]{article}
\usepackage[latin9]{inputenc}
\usepackage{a4wide}
\usepackage{latexsym}
\usepackage{epsf}
\usepackage{amssymb}
\usepackage{ifpdf}
\ifpdf
\usepackage[pdftex]{graphicx}
\usepackage[pdftex,unicode,implicit]{hyperref}
\hypersetup{%
  pdftitle    = {Finding the supersymmetric solutions of N=2 D=4 sugra coupled to vector-multiplets},
  pdfkeywords = {N=2 Supersymmetry, special Kähler geometry, BPS solutions},
  pdfauthor   = {P. Meessen, T. Ortín},
  pdfcreator  = {pdf\LaTeXe\ with package \flqq hyperref\frqq},
  pdfproducer = {pdf\LaTeXe\ with package \flqq hyperref\frqq},
  pdfpagemode = None,  
  pdffitwindow= true,  
  unicode     = true,
  plainpages  = false,
  colorlinks  = true,  
  citecolor   = black,
  urlcolor    = black,
  linkcolor   = black
}
\newcommand{\hepth}[1]{arXiv:{\tt \href{http://www.arXiv.org/abs/hep-th/#1}{hep-th/#1}}}

\else
  \usepackage[dvips]{graphicx}
  \usepackage[unicode,implicit]{hyperref}
  \newcommand{\hepth}[1]{arXiv:{\tt hep-th/#1}}

\fi
   \makeatletter
\@addtoreset{equation}{section}
\makeatother

\begin{document}
\begin{flushright}
\small
IFT-UAM/CSIC-05-50\\
CERN-PH-TH/2005-248\\
{\bf hep-th/0603099}\\
March $13^{th}$, $2006$
\normalsize
\end{flushright}
\begin{center}
\vspace{2cm}
{\Large {\bf The supersymmetric configurations of\\[.5cm] 
$N=2,d=4$ supergravity\\[.5cm]
 coupled to vector supermultiplets}} \vspace{2cm}

{\sl\large Patrick Meessen}${}^{\diamondsuit}$
\footnote{E-mail: {\tt Patrick.Meessen@cern.ch}},
{\sl\large and Tom{\'a}s Ort\'{\i}n}${}^{\spadesuit}$
\footnote{E-mail: {\tt Tomas.Ortin@cern.ch}}

\vspace{1cm}

${}^{\diamondsuit}$\ {\it Physics Department, Theory, CERN, Geneva, 
Switzerland}

\vspace{.2cm}
${}^{\spadesuit}$\ {\it Instituto de F\'{\i}sica Te\'orica UAM/CSIC\\
  Facultad de Ciencias C-XVI,  C.U.~Cantoblanco,  E-28049-Madrid, Spain}\\

\vspace{2cm}


{\bf Abstract}

\end{center}

\begin{quotation}

\small

We classify all the supersymmetric configurations of ungauged
$N=2,d=4$ supergravity coupled to $n$ vector multiplets and determine
under which conditions they are also classical solutions of the
equations of motion. The supersymmetric configurations fall into two
classes, depending on the timelike or null nature of the Killing
vector constructed from Killing spinor bilinears. The timelike class
configurations are essentially the ones found by Behrndt, L\"ust and
Sabra, which exhaust this class and are the ones that include
supersymmetric black holes. The null class configurations include
$pp$-waves and cosmic strings.

\end{quotation}

\newpage

\pagestyle{plain}


\tableofcontents

\newpage

\section{Introduction and main results}

Classical supersymmetric solutions of supergravity theories play a key
rôle in many of the recent developments in string theory, provide
vacua, on which the theory can be quantized and may be interesting for
phenomenology, and objects that live in those vacua such as
$p$-branes, black holes etc.  Therefore, the amount of effort that is
being devoted to the classification of supersymmetric solutions of
supergravity theories, both in higher
\cite{Gauntlett:2002fz}-\cite{Behrndt:2005im} and lower
\cite{Gibbons:1982fy}-\cite{Bellorin:2005zc} dimensions, can hardly be
called a surprise.

$N=2,d=4$ supergravities, the cases under consideration, 
are particularly interesting theories: they are simple enough to
be manageable and yet rich enough in structure, duality symmetries,
interesting solutions and phenomena. Many, but not all, of them are
also related to low-energy limit of Calabi-Yau
compactifications of 10-dimensional type~II superstring theories.
There is a very extensive literature on these theories\footnote{A good
  review on these theories is Ref.~\cite{kn:toinereview}.} but, except
for the simplest cases of pure gauged and ungauged supergravity
\cite{Tod:1983pm,Caldarelli:2003pb}, and for black-hole type
solutions,\footnote{See, {\em e.g.\/} the reviews
  \cite{Mohaupt:2000mj,Andrianopoli:2002fj,deWit:2005ya,Mohaupt:2005jd}
  and references therein.} there have been no systematic attempts to
classify all their supersymmetric solutions. In this work we start
filling this gap by classifying all the supersymmetric configurations
and solutions in the next-to-simplest case, namely pure supergravity
coupled to $n$ vector supermultiplets whose supersymmetric black-hole
solutions have been studied intensively in the not so remote past. 
This work should also lay the
groundwork for the more complicated cases we intend to study next.  

In this work we use the method of Ref.~\cite{Gauntlett:2002nw},
consisting in finding differential and algebraic equations satisfied
by the tensors that can be built as bilinears of the Killing spinor,
whose existence we assume from the onset. We then derive
consistency conditions for these equations to admit solutions and
determine necessary conditions for the backgrounds to be
supersymmetric. Subsequently we show that the conditions are also
sufficient, meaning that we have identified all the supersymmetric
\textit{configurations} of the theory. Finally we impose the equations
of motion in order to find the supersymmetric \textit{solutions}. Throughout this
work we stress the difference between generic supersymmetric field
\textit{configurations} and classical \textit{solutions} of the
equations of motion. We will also make use of the \textit{Killing spinor
  identities}, derived in Refs.~\cite{Kallosh:1993wx,Bellorin:2005hy},
to minimize the number of independent equations of motion that need
to be checked explicitly in order to prove that a given supersymmetric configuration
is a solution.

Let us briefly describe our results: the supersymmetric solutions
of $N=2,d=4$ supergravity coupled to $n$ vector supermultiplets belong
to two main classes: 

\begin{enumerate}
\item Those with a timelike Killing vector. They are essentially the
  field configurations found in Ref.~\cite{Behrndt:1997ny}. These
  solutions were shown in Ref.~\cite{LopesCardoso:2000qm} to be the
  only supersymmetric ones with Killing spinors satisfying the
  constraint Eq.~(\ref{eq:constraint}). Here we show that all the
  supersymmetric solutions in the timelike class admit Killing spinors
  that satisfy that constraint and, therefore there are no more
  supersymmetric configurations nor solutions in this class.
  
  These supersymmetric configurations are completely determined by a
  choice of symplectic section $\mathcal{V}/X$. The metric is then
  given by

\begin{equation}
ds^{2} = |M|^{2}(dt+\omega)^{2} -|M|^{-2}dx^{i}dx^{i}\, ,
\end{equation}

\noindent
where 

\begin{equation}
|M|^{-2} = 2e^{\mathcal{K}[\mathcal{V}/X,(\mathcal{V}/X)^{*}]}\, ,
\end{equation}

\noindent
and where $\mathcal{K}[\mathcal{V}/X,(\mathcal{V}/X)^{*}]$ means that
the K\"ahler potential has to be computed using in the expression
Eq.~(\ref{eq:kpotential}) the components of the symplectic section
$\mathcal{V}/X$.  $\omega=\omega_{\underline{i}}dx^{i}$ is a
time-independent 1-form that has to satisfy the constraint

\begin{equation}
(d\omega)_{mn} =\epsilon_{mnp}e^{-\mathcal{K}
[\mathcal{V}/X,(\mathcal{V}/X)^{*}]} 
\mathcal{Q}_{p}[\mathcal{V}/X,(\mathcal{V}/X)^{*}]\, ,  
\end{equation}

\noindent
where $\mathcal{Q}_{p}[\mathcal{V}/X,(\mathcal{V}/X)^{*}]$ is the
pullback of the K\"ahler 1-form connection, computed in the same fashion.

The vector field strengths are given by

\begin{equation}
F = {\textstyle\frac{1}{\sqrt{2}}} \{d[|M|^{2}\mathcal{R}(dt+\omega) ] 
-{}^{\star}[ |M|^{2} d\mathcal{I}\wedge(dt+\omega)] \}\, .
\end{equation}

\noindent
where $\mathcal{R}$ and $\mathcal{I}$ stand, respectively, for the
real and imaginary parts of the symplectic section $\mathcal{V}/X$.

The scalar fields $Z^{i}$ can be computed by taking the quotients

\begin{equation}
Z^{i}=(\mathcal{V}/X)^{i}/(\mathcal{V}/X)^{0}\, .
\end{equation}

The supersymmetric configurations are classical solutions iff the real
section $\mathcal{I}$ is harmonic on $\mathbb{R}^{3}$. Rewriting the
equation that determines $\omega$ as

\begin{equation}
(d\omega)_{mn} =2\epsilon_{mnp}
\langle\,\mathcal{I}\mid \partial_{p}\mathcal{I}\, \rangle\, ,  
\end{equation}

\noindent
we see that its integrability condition 

\begin{equation}
\langle\,\mathcal{I}\mid \partial_{p}\partial_{p}\mathcal{I}\, \rangle\,  =0\, ,
\end{equation}

\noindent
is satisfied for $\mathcal{I}$ harmonic on $\mathbb{R}^{3}$. In
practice, though, the only functions globally harmonic in
$\mathbb{R}^{3}$ are constant and the rest have singularities and for
them the above condition becomes non-trivial to satisfy
\cite{Denef:2000nb,Bates:2003vx}. We will discuss these conditions
and their implications in a forthcoming paper \cite{kn:BMO}.

The $2\bar{n}$ real harmonic functions then determine the
solution, although one has to solve $\mathcal{R}$ in terms of the
$\mathcal{I}$ in order to be able to write the whole
solution explicitly in terms of the harmonic functions. This problem is
equivalent to that of solving the \textit{stabilization equations} and
has no known generic solution except in a few cases, some of which we review in
Appendix~\ref{appsec:SpecShit}.

\item Those with a null Killing vector \cite{Greene:1989ya}:
  Generically they have Brinkmann-type metrics

\begin{equation}
ds^{2} = 2 du (dv + H du +\hat{\omega})
-2e^{-\mathcal{K}}dzdz^{*}\, .
\end{equation}
  
\noindent
where $\mathcal{K}$ is the K\"ahler potential and $\hat{\omega}$ is
determined by the equation

\begin{equation}
(d\hat{\omega})_{\underline{z}\underline{z}^{*}}
=
2i e^{-\mathcal{K}}\mathcal{Q}_{\underline{u}}\, ,
\end{equation}

\noindent
where $\mathcal{Q}_{\mu}$ is the pullback of the Ka\"hler 1-form
connection (See Eq. (\ref{eq:KahlerConPB})).

The scalar fields can be defined through a symplectic section with
arbitrary dependence on $u$ and $z$ and the vector fields are
determined by complex arbitrary functions
$\phi(u),\psi^{i}(z,z^{*},u)$

\begin{equation}
F =e^{-\mathcal{K}/2} \left(\mathcal{U}_{i}\psi^{i}
+\textstyle{\frac{i}{2}}\mathcal{V}^{*}\phi \right)du\wedge dz^{*} 
+\textrm{c.c.}\, .  
\end{equation}

The solutions of this case are harder to determine completely. There
are, however, two interesting families of solutions:

  \begin{enumerate}
  \item Cosmic strings. They have vanishing vector field strengths
    and scalars that are arbitrary holomorphic functions $Z^{i}(z)$.

\begin{equation}
\left\{
  \begin{array}{rcl}
ds^{2} & = & 2du(dv+Hdu) -2e^{-[\mathcal{K}(Z,Z^{*})-h-h^{*}]} dzdz^{*}\, ,\\
& & \\
Z^{i} & = & Z^{i}(z)\, ,\\
& & \\
h & = & h(z)\, ,\\
& & \\
\partial_{\underline{z}}\partial_{\underline{z}^{*}}H & = & 0\, .\\
  \end{array}
\right.
\end{equation}

The functions $h$ must have the right behavior under K\"ahler
transformations to make the metric formally duality-invariant and the
Killing spinors well defined. These solutions generalize the ones
found in Ref.~\cite{Greene:1989ya} in flat spacetime for arbitrary
K\"ahler potentials. Observe that the harmonic function $H$ describes
a plane wave moving along the string.

\item Plane waves. In the simplest case they have the form

\begin{equation}
\left\{
  \begin{array}{rcl}
ds^{2} & = & 2du(dv+Hdu) -2dzdz^{*}\, ,\\
& & \\
F^{\Lambda\, +} & = &  
\frac{i}{2}\mathcal{L}^{*\, \Lambda}\phi(u) du\wedge dz^{*}\, ,\\
& & \\
Z^{i} & = & Z^{i}(u)\, ,\\
& & \\
H & = & (\mathcal{G}_{ij^{*}}\dot{Z}^{i}\dot{Z}^{*\, j^{*}} +2|\phi|^{2})|z|^{2} 
+f(z,u) +f^{*}(z^{*},u)\, ,\\
  \end{array}
\right.
\end{equation}

\noindent
where $Z^{i},\phi$ are arbitrary functions of $u$ and $f$ an arbitrary
function of $u$ and $z$.

  \end{enumerate}

\end{enumerate}

This work is organized as follows: in section~\ref{sec-n2d4} we review
the aspects of these theories relevant for this work: action,
equations of motion, supersymmetry transformations and symplectic
transformations. In section~\ref{sec-setup} we set up the problem we
want to solve: Killing spinor equations, integrability conditions and
conditions imposed on Killing spinor bilinears. In
section~\ref{sec-timelike} we solve the case in which the Killing
vector bilinear is timelike and in section~\ref{sec-null} the case in
which it is null. The appendices contain the conventions
(\ref{sec-conventions}) some formulae of K\"ahler
(\ref{sec-kahlergeometry}) and special K\"ahler
(\ref{sec-specialgeometry}) geometry plus some explicit examples of
supersymmetric solutions for chosen theories (\ref{appsec:SpecShit}).


\section{$N=2,d=4$ supergravity coupled to vector supermultiplets}
\label{sec-n2d4}

In this section we are going to describe briefly the theory we are
going to work with.  Our main source for the formalism used in this
section is Ref.~\cite{Andrianopoli:1996cm}, whose notation we use here
quite closely although its origin goes back to
Refs.~\cite{deWit:1984pk,deWit:1984px}. Our conventions for the
metric, connection, curvature, gamma matrices and spinors are
described in detail in the appendices of Ref.~\cite{Bellorin:2005zc}
which also contain many identities and results that will be used
repeatedly throughout the text.  These conventions are very similar,
but not identical, to those employed in
Ref.~\cite{Andrianopoli:1996cm}.  The differences and a dictionary of
all the indices we use can be found in Appendix~\ref{sec-conventions}.

The gravity multiplet of the $N=2,d=4$ theory consists of the
graviton $e^{a}{}_{\mu}$, a pair of gravitinos $\psi_{I\, \mu}\,
,\,\,\, (I=1,2)$ which we describe as Weyl spinors, and a vector field
$A_{\mu}$.  Each of the $n$ vector supermultiplets of $N=2,d=4$
supergravity that we are going to couple to the pure supergravity
theory contains complex scalar $Z^{i}\, ,\,\,\,\, (i=1,\cdots, n)$, a
pair of gauginos $\lambda^{I\, i}$, which we also describe as Weyl
spinors and a vector field $A^{i}{}_{\mu}$. In the coupled theory, the
$\bar{n}=n+1$ vectors can be treated on the same footing and they are
described collectively by an array $A^{\Lambda}{}_{\mu}\,\,\,\,\,
(\Lambda=1,\cdots,\bar{n})$. The coupling of scalars to scalars is
described by a non-linear $\sigma$-model with K\"ahler metric
$\mathcal{G}_{ij^{*}}(Z,Z^{*})$ (see
Appendix~\ref{sec-kahlergeometry}), and the coupling to the vector
fields by a complex scalar-field-valued matrix
$\mathcal{N}_{\Lambda\Sigma}(Z,Z^{*})$.  These two couplings are
related by a structure called special K\"ahler geometry, described
in Appendix~\ref{sec-specialgeometry}. The symmetries of these two
sectors will be related and this relation will be discussed shortly.

The action for the bosonic fields of the theory is

\begin{equation}
\label{eq:action}
 S= \int d^{4}x \sqrt{|g|}
\left[R +2\mathcal{G}_{ij^{*}}\partial_{\mu}Z^{i}
\partial^{\mu}Z^{*\, j^{*}}
+2\Im{\rm m}\mathcal{N}_{\Lambda\Sigma} 
F^{\Lambda\, \mu\nu}F^{\Sigma}{}_{\mu\nu}
-2\Re{\rm e}\mathcal{N}_{\Lambda\Sigma} 
F^{\Lambda\, \mu\nu}{}^{\star}F^{\Sigma}{}_{\mu\nu}
\right]\, .
\end{equation}

\noindent
Observe that the canonical normalization of the vector fields kinetic
terms implies that $\Im{\rm m}\mathcal{N}_{\Lambda\Sigma}$ is negative
definite, as is guaranteed by special geometry \cite{Craps:1997gp}.

For vanishing fermions, the supersymmetry transformation rules of the
fermions are

\begin{eqnarray}
\delta_{\epsilon}\psi_{I\, \mu} & = & 
\mathfrak{D}_{\mu}\epsilon_{I} 
+\epsilon_{IJ}T^{+}{}_{\mu\nu}\gamma^{\nu}\epsilon^{J}\, ,
\label{eq:gravisusyrule}\\
& & \nonumber \\
\delta_{\epsilon}\lambda^{Ii} & = & 
i\not\!\partial Z^{i}\epsilon^{I} +\epsilon^{IJ}\not\!G^{i\, +}\epsilon_{J}\, ,
\label{eq:gaugsusyrule}
\end{eqnarray}

\noindent
where $\mathfrak{D}_{\mu}$ is defined in Eq.~(\ref{eq:Kcovariantderivative2}), 
which acts on the spinors $\epsilon_{I}$, since they are of K\"ahler weight $1/2$, as

\begin{equation}
\mathfrak{D}_{\mu}\epsilon_{I} \equiv 
(\nabla_{\mu} +{\textstyle\frac{i}{2}}\mathcal{Q}_{\mu})\epsilon_{I}\, ,
\end{equation}

\noindent
and $\mathcal{Q}_{\mu}$ is the pullback of the K\"ahler 1-form defined
in Eq.~(\ref{eq:K1form}). The 2-forms $T$ and $G^{i}$ are the
combinations

\begin{eqnarray}
T_{\mu\nu} & \equiv & \mathcal{T}_{\Lambda}F^{\Lambda}{}_{\mu\nu}\, ,\\
& & \nonumber \\
G^{i}{}_{\mu\nu} & \equiv &
\mathcal{T}^{i}{}_{\Lambda}F^{\Lambda}{}_{\mu\nu}\, ,
\end{eqnarray}

\noindent
where, in turn, $\mathcal{T}_{\Lambda}$ and $\mathcal{T}^{i}{}_{\Lambda}$ are,
respectively, the graviphoton and the matter vector fields projectors, defined
in Eqs.~(\ref{eq:projectorg}) and (\ref{eq:projectorm}).

The supersymmetry transformations of the bosons are

\begin{eqnarray}
  \delta_{\epsilon} e^{a}{}_{\mu} & = & 
-{\textstyle\frac{i}{4}} (\bar{\psi}_{I\, \mu}\gamma^{a}\epsilon^{I}
+\bar{\psi}^{I}{}_{\mu}\gamma^{a}\epsilon_{I})\, ,
\label{eq:susytranse}\\
& & \nonumber \\ 
  \delta_{\epsilon} A^{\Lambda}{}_{\mu} & = & 
{\textstyle\frac{1}{4}}
(\mathcal{L}^{\Lambda\, *}
\epsilon^{IJ}\bar{\psi}_{I\, \mu}\epsilon_{J}
+
\mathcal{L}^{\Lambda}
\epsilon_{IJ}\bar{\psi}^{I}{}_{\mu}\epsilon^{J}) \nonumber \\
& & \nonumber \\
& & 
+
{\textstyle\frac{i}{8}}(f^{\Lambda}{}_{i}\epsilon_{IJ}
\bar{\lambda}^{Ii}\gamma_{\mu}
\epsilon^{J}
+
f^{\Lambda *}{}_{i^{*}}\epsilon^{IJ}
\bar{\lambda}_{I}{}^{i^{*}}\gamma_{\mu}\epsilon_{J})\, ,
\label{eq:susytransA}\\
& & \nonumber \\
  \delta_{\epsilon} Z^{i} & = & 
{\textstyle\frac{1}{4}} \bar{\lambda}^{Ii}\epsilon_{I}\, .
\label{eq:susytransZ}
\end{eqnarray}

For convenience, we denote  the Bianchi identities for the vector field strengths by

\begin{equation}
\label{eq:BL}
\mathcal{B}^{\Lambda\, \mu} \equiv \nabla_{\nu}{}^{\star}F^{\Lambda\,
  \nu\mu}\, .  
\end{equation}

\noindent
and the bosonic equations of motion by

\begin{equation}
\mathcal{E}_{a}{}^{\mu}\equiv 
-\frac{1}{2\sqrt{|g|}}\frac{\delta S}{\delta e^{a}{}_{\mu}}\, ,
\hspace{.5cm}
\mathcal{E}_{i} \equiv -\frac{1}{2\sqrt{|g|}}
\frac{\delta S}{\delta Z^{i}}\, ,
\hspace{.5cm}
\mathcal{E}_{\Lambda}{}^{\mu}\equiv 
\frac{1}{8\sqrt{|g|}}\frac{\delta S}{\delta A^{\Lambda}{}_{\mu}}\, ,
\end{equation}

whose explicit forms can be found to be

\begin{eqnarray}
\mathcal{E}_{\mu\nu} & = & 
G_{\mu\nu}
+2\mathcal{G}_{ij^{*}}[\partial_{\mu}Z^{i} \partial_{\nu}Z^{*\, j^{*}}
-{\textstyle\frac{1}{2}}g_{\mu\nu}
\partial_{\rho}Z^{i}\partial^{\rho}Z^{*\, j^{*}}]\nonumber \\
& & \nonumber \\
& & 
+8\Im {\rm m}\mathcal{N}_{\Lambda\Sigma}
F^{\Lambda\, +}{}_{\mu}{}^{\rho}F^{\Sigma\, -}{}_{\nu\rho}\, ,
\label{eq:Emn}\\
& & \nonumber \\
\mathcal{E}_{i} & = & \nabla_{\mu}(\mathcal{G}_{ij^{*}}
\partial^{\mu}Z^{*\, i^{*}})
-\partial_{i}\mathcal{G}_{jk^{*}}\partial_{\rho}Z^{j}
\partial^{\rho}Z^{*\, k^{*}} 
+\partial_{i}[
\tilde{F}_{\Lambda}{}^{\mu\nu}{}^{\star}F^{\Lambda}{}_{\mu\nu}]\, ,
\label{eq:Ei}\\
& & \nonumber \\
\mathcal{E}_{\Lambda}{}^{\mu} & = & 
\nabla_{\nu}{}^{\star}\tilde{F}_{\Lambda}{}^{\nu\mu}\, ,
\label{eq:ERm}
\end{eqnarray}

\noindent
where we have defined the dual vector field strength $\tilde{F}_{\Lambda}$
by 

\begin{equation}
\tilde{F}_{\Lambda\, \mu\nu} \equiv   
-\frac{1}{4\sqrt{|g|}}\frac{\delta S}{\delta {}^{\star}F^{\Lambda}{}_{\mu\nu}}
= \Re {\rm e}\mathcal{N}_{\Lambda\Sigma}F^{\Sigma}{}_{\mu\nu}
+\Im {\rm m}\mathcal{N}_{\Lambda\Sigma}{}^{*}F^{\Sigma}{}_{\mu\nu}\, .
\end{equation}

The Maxwell and Bianchi identities can be rotated into each other by
$GL(2\bar{n},\mathbb{R})$ transformations under which they are a
$2\bar{n}$-dimensional vector:

\begin{equation}
\label{eq:maxwellvector}
\mathcal{E}^{\mu}\equiv \left(
  \begin{array}{c}
\mathcal{B}^{\Lambda\, \mu}\\
\\
\mathcal{E}_{\Lambda}{}^{\mu}\\
  \end{array}
\right)  
\longrightarrow
\left(
  \begin{array}{cc}
D & C \\
& \\
B & A \\
  \end{array}
\right)
\left(
  \begin{array}{c}
\mathcal{B}^{\Lambda\, \mu}\\
\\

\mathcal{E}_{\Lambda}{}^{\mu}\\
  \end{array}
\right)\, ,
\end{equation}

\noindent
where $A,B,C$ and $D$ are $\bar{n}\times \bar{n}$ matrices. These
transformations act in the same form on the vector of $2\bar{n}$ 2-forms

\begin{equation}
\label{eq:2formvector}
F \equiv 
\left(
  \begin{array}{c}
F^{\Lambda}\\
\\
\tilde{F}_{\Lambda}\\
  \end{array}
\right)  
\longrightarrow
\left(
  \begin{array}{cc}
D & C \\
& \\
B & A \\
  \end{array}
\right)
\left(
  \begin{array}{c}
F^{\Lambda}\\
\\
\tilde{F}_{\Lambda}\\
  \end{array}
\right)\, ,
\end{equation}

\noindent
and, since, by definition,

\begin{equation}
\tilde{F}^{\prime}_{\Lambda} 
= \Re {\rm e}\mathcal{N}^{\prime}_{\Lambda\Sigma}F^{\prime\Sigma}
+\Im {\rm m}\mathcal{N}^{\prime}_{\Lambda\Sigma}{}^{\star}F^{\prime\Sigma}\, ,
\end{equation}

\noindent
for the transformations to be consistently defined, the must act on the period
matrix $\mathcal{N}$ according to

\begin{equation}
\mathcal{N}^{\prime} = (A\mathcal{N}+B)(C\mathcal{N}+D)^{-1}\, .  
\end{equation}

\noindent
Furthermore, the transformations must preserve the symmetry of the period
matrix, which requires

\begin{equation}
A^{T}C=C^{T}A\, ,
\hspace{1cm}
D^{T}B=B^{T}D\, ,  
\hspace{1cm}
A^{T}D-C^{T}B=1\, ,
\end{equation}

\noindent
{\em i.e.\/} the transformations must belong to $Sp(2\bar{n},\mathbb{R})$.

The above transformation rules for the vector field strength and period matrix
imply

\begin{equation}
\Im{\rm m} \mathcal{N}^{\prime} =(C\mathcal{N}^{*}+D)^{-1\, T}   
\Im{\rm m} \mathcal{N} (C\mathcal{N}+D)^{-1}\, ,
\hspace{1cm}
F^{\prime\Lambda\, +} = (C\mathcal{N}^{*}+D)_{\Lambda\Sigma}F^{\Sigma\, +}\, ,
\end{equation}

\noindent
so the combination $\Im {\rm m}\mathcal{N}_{\Lambda\Sigma} F^{\Lambda\,
  +}{}_{\mu}{}^{\rho}F^{\Lambda\, -}{}_{\nu\rho}$ that appears in the
energy-momentum tensor is automatically invariant.

The above symplectic transformations of the period matrix $\mathcal{N}$
correspond to certain transformations of the complex scalar fields $Z^{i}$:

\begin{equation}
\mathcal{N}^{\prime}(Z,Z^{*}) = 
[A\mathcal{N}(Z,Z^{*})+B][C\mathcal{N}(Z,Z^{*})+D]^{-1} \equiv 
\mathcal{N}(Z^{\prime},Z^{\prime\, *})\, .  
\end{equation}

\noindent
These transformations have to be symmetries of the theory as well,
which implies that they have to be isometries of the special K\"ahler
manifold \cite{Gaillard:1981rj}. Thus only the isometries of the
special K\"ahler manifold which are embedded in
$Sp(2\bar{n},\mathbb{R})$ are symmetries of all the equations
of motion of the theory (dualities of the theory). Observe that the
K\"ahler potential is, in general, not invariant under the isometries
of the K\"ahler metric, but undergoes a K\"ahler transformation. This
means that all objects with non-zero K\"ahler weight transform
non-trivially under duality. 

The scalar equation $\mathcal{E}_{i}$ Eq.~(\ref{eq:Ei}) can be written
in a manifestly covariant form by rising the index with
$\mathcal{G}^{j^{*}i}$, $\mathcal{E}^{i^{*}}$. The complex conjugate
equation then takes on the form

\begin{equation}
\label{eq:Ei2}
\mathcal{E}^{i} = \mathfrak{D}_{\mu}\partial^{\mu}Z^{i} 
+\mathcal{G}^{ij^{*}} \partial_{j^{*}}
[\tilde{F}_{\Lambda}{}^{\mu\nu}{}^{*}F^{\Lambda}{}_{\mu\nu}]\, .
\end{equation}


\section{Supersymmetric configurations: general setup}
\label{sec-setup}

Our first goal is to find all the bosonic field configurations
$\{g_{\mu\nu},F^{\Lambda}{}_{\mu\nu}, Z^{i}\}$ for which the Killing
spinor equations (KSEs):

\begin{eqnarray}
\delta_{\epsilon}\psi_{I\, \mu} & = & 
\mathfrak{D}_{\mu}\epsilon_{I} 
+\epsilon_{IJ}T^{+}{}_{\mu\nu}\gamma^{\nu}\epsilon^{J} =0\, ,
\label{eq:KSE1}\\
& & \nonumber \\
\delta_{\epsilon}\lambda^{Ii} & = & 
i\not\!\partial Z^{i}\epsilon^{I} 
+\epsilon^{IJ}\not\!G^{i\, +}\epsilon_{J} =0\, ,
\label{eq:KSE2}
\end{eqnarray}

\noindent
admit at least one solution.\footnote{The generalized holonomy of the
  gravitino supersymmetry transformation
  indicates that the minimal number of solutions
  that these equations will admit is actually 4
  \cite{Batrachenko:2004su}, but we will not use this fact in our
  derivation. We will, in the end recover the result that
  supersymmetric solutions generically preserve $1/2$ or all
  supersymmetries.} It must be stressed that the configurations considered need not be
classical solutions of the equations of motion. Furthermore, we will
not assume that the Bianchi identities are satisfied by the field
strengths of a configuration.

Our second goal will be to identify among all the supersymmetric field
configurations those that satisfy all the equations of motion (including
the Bianchi identities).

Let us initiate the analysis of the KSEs by studying their integrability conditions.


\subsection{Killing Spinor Identities (KSIs)}

Using the supersymmetry transformation rules of the bosonic fields
Eqs.~(\ref{eq:susytranse}--\ref{eq:susytransZ}) and
using the results of Refs.~\cite{Kallosh:1993wx,Bellorin:2005hy} we can derive
the following relations (\textit{Killing spinor identities}, KSIs) between the
(off-shell) equations of motion of the bosonic fields
Eqs.~(\ref{eq:Emn}--\ref{eq:ERm}) that are satisfied by any
field configuration $\{e^{a}{}_{\mu},A^{\Lambda}{}_{\mu},Z^{i}\}$ admitting
Killing spinors:

\begin{eqnarray}
\mathcal{E}_{a}{}^{\mu}\gamma^{a}\epsilon^{I}
-4i\epsilon^{IJ}\mathcal{L}^{\Lambda}
\mathcal{E}_{\Lambda}{}^{\mu} \epsilon_{J}
& = & 0\, , \label{eq:ksi1}\\
& & \nonumber \\
\mathcal{E}^{i}\epsilon^{I} 
-2i\epsilon^{IJ}f^{*\, i\Lambda}\not\!\mathcal{E}_{\Lambda} 
\epsilon_{J} & = & 0\, .
\label{eq:ksi2}
\end{eqnarray}

The vector field Bianchi identities Eq.~(\ref{eq:BL}) do not appear in these
relations because the procedure used to derive them, assumes the existence of
the vector potentials, and hence the vanishing of the Bianchi identities.

It is convenient to treat the Maxwell equations and Bianchi identities on
an equal footing as to preserve the electric-magnetic dualities of the theory, for which
it is convenient to have a duality-covariant version of the above
KSIs. This can be found by performing duality rotations on the above
identities or from the integrability conditions of the KSEs
Eqs.~(\ref{eq:KSE1},\ref{eq:KSE2}), which is the method we are going to use.

Using the K\"ahler special geometry machinery, we obtain

\begin{equation}
 \label{eq:ksi1-1}
  \begin{array}{rcl}
\mathfrak{D}_{[\mu} \delta_{\epsilon} \psi_{I\, \nu]} & = & 
-{\textstyle\frac{1}{8}}\{[R_{\mu\nu}{}^{ab} 
-8\mathcal{T}_{\Lambda}\mathcal{T}^{*}_{\Sigma} 
F^{\Lambda\,  +}{}_{[\mu|}{}^{a} F^{\Sigma\,  -}{}_{|\nu]}{}^{b}]\gamma_{ab}
+4\mathcal{G}_{ij^{*}}\partial_{[\mu}Z^{i}\partial_{\nu]}Z^{*\, j^{*}}
 \}\epsilon_{I} \\
& & \\
& & 
+\epsilon_{IJ}
\mathfrak{D}_{[\mu}T^{+}{}_{\nu]\rho}\gamma^{\rho}\epsilon^{J}=0\, ,\\
\end{array}
\end{equation}

\noindent
which gives rise to

\begin{equation}
 \label{eq:ksi1-2}
4\gamma^{\nu}\mathfrak{D}_{[\mu} \delta_{\epsilon} \psi_{I\, \nu]} =
(\mathcal{E}_{\mu\nu} 
-{\textstyle\frac{1}{2}}g_{\mu\nu}\, \mathcal{E}_{\sigma}{}^{\sigma})
\gamma^{\nu}\epsilon_{I} 
-2i \epsilon_{IJ}\mathcal{L}^{\Lambda}
(\not\! \mathcal{E}_{\Lambda}  
-\mathcal{N}_{\Lambda\Sigma}\not\!\!\mathcal{B}^{\Sigma})
\gamma_{\mu}\epsilon^{J}=0\, .
\end{equation}

Contracting the above identity with $\gamma^{\mu}$, we obtain another one involving
only the trace $ \mathcal{E}_{\sigma}{}^{\sigma}$, which can be used
to eliminate it completely from the KSIs. The result is the
duality-covariant version of (the complex conjugate of) Eq.~(\ref{eq:ksi1}) we
were after:

\begin{equation}
 \label{eq:ksi1-3}
\mathcal{E}_{a}{}^{\mu}\gamma^{a}\epsilon_{I}
-4i\epsilon_{IJ}\mathcal{L}^{\Lambda}(\mathcal{E}_{\Lambda}{}^{\mu} 
-\mathcal{N}_{\Lambda\Sigma}\mathcal{B}^{\Sigma\mu}) \epsilon^{J}
= 0\, .  
\end{equation}

It turns out to be convenient to define the combination

\begin{equation}
\mathcal{H}^{\Lambda\mu}\equiv 
(\Im{\rm m}\mathcal{N})^{-1|\Lambda\Sigma}(\mathcal{E}_{\Sigma}{}^{\mu} 
-\mathcal{N}_{\Sigma\Omega}\mathcal{B}^{\Sigma\mu})\, .
\end{equation}

\noindent
Using it, the above KSIs Eqs.~(\ref{eq:ksi1-2},\ref{eq:ksi1-3}) take the
form

\begin{eqnarray}
 \label{eq:ksi1-4}
(\mathcal{E}_{\mu\nu} 
-{\textstyle\frac{1}{2}}g_{\mu\nu}\, \mathcal{E}_{\sigma}{}^{\sigma})
\gamma^{\nu}\epsilon_{I} 
-\mathcal{T}_{\Lambda}\not\!\!\mathcal{H}^{\Lambda}
\gamma_{\mu}\epsilon_{IJ}\epsilon^{J} & = & 0\, ,\\
& & \nonumber \\
 \label{eq:ksi1-5}
\mathcal{E}_{a}{}^{\mu}\gamma^{a}\epsilon_{I}
-2\mathcal{T}_{\Lambda}\mathcal{H}^{\Lambda\mu}\epsilon_{IJ}\epsilon^{J}
& = & 0\, .  
\end{eqnarray}

Observe that the graviphoton-projected combination
$\mathcal{T}_{\Lambda}\mathcal{H}^{\Lambda\, \mu}$ can be written as

\begin{equation}
\label{eq:localcentralcharge}
\mathcal{T}_{\Lambda}\mathcal{H}^{\Lambda\, \mu} = 
2i [\mathcal{L}^{\Lambda}\mathcal{E}_{\Lambda}{}^{\mu}
-\mathcal{M}_{\Lambda}\mathcal{B}^{\Lambda\, \mu}]
=2i\, \langle\, \mathcal{E}^{\mu}\mid \mathcal{V}\, \rangle\, ,
\end{equation}

\noindent
where $\mathcal{E}$ is the symplectic vector defined in
Eq.~(\ref{eq:maxwellvector}).

The duality-covariant version of Eq.~(\ref{eq:ksi2}) can be obtained in a similar fashion,
and reads

\begin{equation}
 \label{eq:ksi2-1}
-i\not\!\mathfrak{D}\delta_{\epsilon}\lambda^{Ii} =
\mathcal{E}^{i}\epsilon^{I} 
-2i\mathcal{T}^{i}{}_{\Lambda}\not\!\!\mathcal{H}^{\Lambda}
\epsilon^{IJ}\epsilon_{J}
=0\, .  
\end{equation}

Observe that the identities Eqs.~(\ref{eq:ksi1-2},\ref{eq:ksi1-3}) and
(\ref{eq:ksi2-1}) are necessary but not sufficient conditions to have
supersymmetry.

{}From these identities we can derive further identities involving only tensors
by multiplication with gamma matrices and conjugate spinor from the left, as to have
only bilinears. As is usual, it is convenient to consider the case in which the
vector bilinear $V^{\mu}\equiv i\bar{\epsilon}^{I}\gamma^{\mu}\epsilon_{I}$ is
timelike and the case in which it is null, separately.


\subsubsection{The timelike case: independent e.o.m.'s}

When $V^{\mu}$ is timelike one can derive the following identities:

\begin{eqnarray}
\label{eq:ksit1}
\mathcal{E}^{\mu\nu} & = & \mathcal{E}^{\rho\sigma}v_{\rho}v_{\sigma} 
v^{\mu}v^{\nu}\, ,\\
& & \nonumber \\
\label{eq:ksit2}
\mathcal{T}_{\Lambda}\mathcal{H}^{\Lambda\, \mu} & = &   
-{\textstyle\frac{i}{2}}e^{i\alpha}
\mathcal{E}^{\rho\sigma}v_{\rho}v_{\sigma} v^{\mu}\, ,\\
& & \nonumber \\
\label{eq:ksit3}
\mathcal{T}^{i}{}_{\Lambda}\mathcal{H}^{\Lambda\, \mu} & = &   
{\textstyle\frac{1}{2}}e^{-i\alpha}\mathcal{E}^{i}v^{\mu}\, ,
\end{eqnarray}

\noindent
where we have defined the unit vector and the (local) phase 

\begin{equation}
v^{\mu}\equiv V^{\mu}/2|X|\, ,\hspace{1cm} e^{i\alpha}\equiv X/|X|\, .
\end{equation}

These identities contain a large amount of information about the
supersymmetric configurations. In particular, they contain the 
necessary information about which equations of motion need to 
be checked explicitly in order to determine whether a given configuration
solves the equations of motion:
the first of these identities tells us that the only components of the
Einstein equations that do not vanish automatically for supersymmetric
configurations are those in the direction of $v^{\mu}v^{\nu}$; the
rest vanish automatically. {\em I.e.\/} once supersymmetry is
established, one does not need to check that those components of the
Einstein equations are satisfied. Further, the second and third
identities state that the only components of the combination of
Maxwell equations and Bianchi identities $\mathcal{H}^{\Lambda\, \mu}$
that do not vanish automatically are the ones in the direction
$v^{\mu}$.  For the graviphoton (second equation), they are related to
the only non-trivial components of the Einstein equations and for the
matter vector fields (third equation), they are related to the
equations of motion of the scalars. Therefore, we see that iff
the Maxwell equation and Bianchi identities are satisfied, then the equations of
motion of the scalars and the Einstein equations are satisfied identically.
The conclusion then must be that, in the timelike case, one only needs to 
solve the Maxwell equation and the Bianchi identities in order to be sure
that a supersymmetric configuration is an actual (supersymmetric) solution of 
the equations of motion. 

\subsubsection{The null case}

When $V^{\mu}$ is a null-vector (we will denote it by $l^{\mu}$), using the
auxiliary spinor $\eta$ defined in the appendix of
Ref.~\cite{Bellorin:2005zc} to construct a standard complex null
tetrad $\{l^{\mu},n^{\mu},m^{\mu},m^{*\, \mu}\}$ we can derive the
following identities:

\begin{eqnarray}
(\mathcal{E}_{\mu\nu} 
-{\textstyle\frac{1}{2}}g_{\mu\nu}\, \mathcal{E}_{\sigma}{}^{\sigma}) l^{\nu}  
=
(\mathcal{E}_{\mu\nu} 
-{\textstyle\frac{1}{2}}g_{\mu\nu}\, \mathcal{E}_{\sigma}{}^{\sigma}) m^{\nu}
& = & 0\, ,
\label{eq:ksinull1}\\
& & \nonumber \\
\mathcal{E}_{\mu\nu} l^{\nu}  
=
\mathcal{E}_{\mu\nu} m^{\nu}
& = & 0\, ,\\
& & \nonumber \\
\mathcal{T}_{\Lambda}\mathcal{H}^{\Lambda\, \mu} & = & 0\, ,
\label{eq:ksinull3}\\
& & \nonumber \\
\mathcal{T}^{i}{}_{\Lambda}\mathcal{H}^{\Lambda\, \mu} l_{\mu}
= 
\mathcal{T}^{i}{}_{\Lambda}\mathcal{H}^{\Lambda\, \mu} m_{\mu}
& = & 0\, ,
\label{eq:ksinull4}\\
& & \nonumber \\
\mathcal{E}^{i} & = & 0\, .
\label{eq:ksinull5}
\end{eqnarray}

Thus, in this case, the equations of motion of the scalars are always
automatically satisfied for a supersymmetric configuration. Only a few
components of the Einstein and Maxwell equations and Bianchi
identities may also be non-zero and these are the only ones that need
to be checked if we want to have solutions. Observe that the
vanishing of the graviphoton-projected combination
$\mathcal{T}_{\Lambda}\mathcal{H}^{\Lambda\, \mu}$ does not imply the
vanishing of the Maxwell equations or the Bianchi identities.


\subsection{Solving the Killing spinor equations}
\label{sec-solving}

To solve the KSEs we are going to follow these steps:

\begin{enumerate}
\item In section~\ref{sec-bilinearkse}, we are going to derive equations for
  the tensor bilinears that can be built from the Killing
  spinors.\footnote{The definitions and properties of these bilinears can be
    found in the appendices of Ref.~\cite{Bellorin:2005zc}.} Solving these
  equations is not, in principle, sufficient for solving the KSEs, but it is
  certainly necessary, which is why they are analyzed first.
\item We are going to see in in the same section that these equations
  for the bilinears state that the vector bilinear we denote by
  $V^{\mu}$, is always a Killing vector, whereas the other three are closed
  (locally exact) 1-forms, which need not be independent.
\item In the same section we will derive an expression for the contractions
  $V^{\nu} F^{\Lambda}{}_{\nu\mu}$ in terms of the scalar bilinear $X$ and the
  scalars $Z^{i}$. These contractions determine to a large extent the form of
  the full vector field strengths, depending on the causal nature of the
  Killing vector $V^{\mu}$, which can be timelike or null. These two cases
  have to be studied separately.
\item In the timelike case (section~\ref{sec-timelike}) 
  \begin{enumerate}
  \item The contractions $V^{\nu} F^{\Lambda}{}_{\nu\mu}$ fully determine the
    vector field strengths (section~\ref{sec-vectorfieldstrengths}).  
  \item The form of the metric is also fixed by the existence of a timelike
    Killing vector and the three exact 1-forms, which, in this case, are independent
    (section~\ref{sec-metric}).
  \item At this point these two fields are entirely expressed in terms of bilinear
    $X$ and the scalars $Z^{i}$, which remain arbitrary, and we are going to
    check explicitly (section~\ref{sec-solvingKSEtimelike}) that, in all
    cases, these field configurations are supersymmetric,
    provided that they satisfy the
    integrability condition Eq.~(\ref{eq:dointegrability}).
    
  \item This solves the timelike case, but, obviously, we are
    particularly interested in supersymmetric configurations which are
    \textit{solutions}.  We have seen in the previous section that the
    KSIs insure that this is equivalent to satisfying the Maxwell equations and
    the Bianchi identities, which, as we are going to see (section~\ref{sec-eoms}),
    is the case if the scalars satisfy the `simple' Eqs.~(\ref{eq:laplaceI}).

  \end{enumerate}
  
\item In the null case (section~\ref{sec-null})

  \begin{enumerate}
  \item we will use the formalism of Ref.~\cite{Tod:1995jf},
    exploiting the fact that the two $\epsilon_{I}$ must be
    proportional and can be written in the form
    $\epsilon_{I}=\phi_{I}\epsilon$. The KSEs can be split into
    equations involving $\epsilon$ and equations involving
    $\phi_{I}$s.
    
  \item A second spinor $\eta$ needs to be introduced as to construct a
    null tetrad via spinor bilinears; the relative normalization of
    $\epsilon$ and $\eta$ requires $\eta$ to satisfy a differential
    equation whose integrability conditions need to be added to the
    KSEs integrability conditions. All these conditions and the
    conditions implied for the null tetrad are studied in
    section~\ref{eq:eomintegrabilityconstraintsnull}.

  \item Since the solution admits a covariantly constant null vector,
        we can introduce a coordinate system and solve the consistency
        conditions (see section~\ref{sec:NullMetric}). In section~\ref{sec:NullKSE}
        we use this coordinate system to analyze the KSEs, and show that
        a supersymmetric configuration preserves either half or all the
        supersymmetries.

  \item Section~\ref{sec:NullEOM}, then, analyzes the equations of motion,
        reducing them to two, seemingly involved, differential equations, 
        namely Eqs. (\ref{eq:QueHorror}) and (\ref{eq:EEuu}), and discusses
        some interesting subclasses of solutions.     

\end{enumerate}

\end{enumerate}


\subsection{Killing equations for the bilinears}
\label{sec-bilinearkse}

From the gravitino supersymmetry transformation rule
Eq.~(\ref{eq:gravisusyrule}) we get the independent equations

\begin{eqnarray}
\mathfrak{D}_{\mu}X & = & -i T^{+}{}_{\mu\nu}V^{\nu}  \, ,
\label{eq:TV}\\
& & \nonumber \\
\nabla_{\mu} V^{I}{}_{J\, \nu} & = & 
i \delta^{I}{}_{J} (XT^{*-}{}_{\mu\nu} -X^{*}T^{+}{}_{\mu\nu})
-i(\epsilon^{IK}T^{*\, -}{}_{\mu\rho}\Phi_{KJ}{}^{\rho}{}_{\nu}
-\epsilon_{JK}T^{+}{}_{\mu\rho}\Phi^{IK}{}_{\nu}{}^{\rho})\, .
\end{eqnarray}

\noindent
The first equation relates the scalar bilinear $X$ with the self-dual
part of the graviphoton field strength and indicates that it contains
all the information of the central charge of the theory
\cite{Ceresole:1995jg}.

The first term in the r.h.s.~of the second equation is completely
antisymmetric in $\mu\nu$ indices and has a non-vanishing trace in
$IJ$ indices, while the second term is completely symmetric in
$\mu\nu$ indices and traceless in $IJ$ indices.  This implies that
$V^{\mu}$ is a Killing vector and the 1-form $\hat{V}=V_{\mu}dx^{\mu}$
satisfies the equation

\begin{equation}
\label{eq:dV}
d\hat{V} = 4i (XT^{*\, -} -X^{*}T^{+})\, ,
\end{equation}

\noindent
while the remaining 3 independent 1-forms $\hat{V}^{i}\equiv
\frac{1}{\sqrt{2}} V^{I}{}_{J\, \mu}\sigma^{i\, J}{}_{I} dx^{\mu}$
($\sigma^{i\, I}{}_{J}\, ,\,\,\, i=1,2,3$ are the Pauli matrices) are exact

\begin{equation}
d\hat{V}^{i} =0\, .
\end{equation}

From the gauginos supersymmetry transformation rules,
Eqs.~(\ref{eq:gaugsusyrule}), we get 

\begin{eqnarray}
V^{I}{}_{K}{}^{\mu}\partial_{\mu}Z^{i} +\epsilon^{IJ}\Phi_{KJ}{}^{\mu\nu}
G^{i\, +}{}_{\mu\nu} & = & 0\, ,\\
& & \nonumber \\
iM^{KI}\partial_{\mu}Z^{i} +i\Phi^{KI}{}_{\mu}{}^{\nu}\partial_{\nu}Z^{i}
-4i\epsilon^{IJ}V^{K}{}_{J}{}^{j}G^{i\, +}{}_{\mu\nu}  & = & 0\, .
\end{eqnarray}

\noindent
The trace of the first equation gives

\begin{equation}
  \label{eq:VdZ}
V^{\mu}\partial_{\mu}Z^{i} =0\, ,  
\end{equation}

\noindent
while the antisymmetric part of the second equation gives

\begin{equation}
\label{eq:GV}
2i X^{*}\partial_{\mu}Z^{i}+4i G^{i\, +}{}_{\mu\nu}V^{\nu} =0\, .  
\end{equation}

\noindent
Using Eq. (\ref{eq:SGImpId}), we can derive

\begin{equation}
\label{eq:completeness}
i\mathcal{L}^{*\, \Lambda}T^{+} 
+2f^{\Lambda}{}_{i}G^{i\, +}= F^{\Lambda\, +}\, ,  
\end{equation}

\noindent
which in its turn allows us to combine Eqs.~(\ref{eq:TV}) and (\ref{eq:GV}), as to obtain

\begin{equation}
\label{eq:FV}
V^{\nu}F^{\Lambda\, +}{}_{\nu\mu}= 
\mathcal{L}^{*\, \Lambda}\mathfrak{D}_{\mu}X
+X^{*} f^{\Lambda}{}_{i}\partial_{\mu}Z^{i} =
 \mathcal{L}^{*\, \Lambda}\mathfrak{D}_{\mu}X
+X^{*}\mathfrak{D}_{\mu} \mathcal{L}^{\Lambda}\, ,
\end{equation}

\noindent
which, in the timelike case, is enough to completely determine
$F^{\Lambda}$ as a function of the scalars $Z^{i},X$ and $V$.


\section{The timelike case}
\label{sec-timelike}


\subsection{The vector field strengths}
\label{sec-vectorfieldstrengths}

As is well-known, the contraction of a self-dual 2-form with a non-null vector completely
determines the 2-form. In the timelike case we can use $V^{\mu}$ and we have

\begin{equation}
\label{eq:decomposition2}
C^{\Lambda\, +}{}_{\mu}\equiv V^{\nu}
F^{\Lambda\, +}{}_{\nu\mu}\,\,\, \Rightarrow\,\,\,
F^{\Lambda\, +}=V^{-2}[\hat{V}\wedge \hat{C}^{\Lambda\, +}
+ i\,{}^{\star}\!(\hat{V}\wedge \hat{C}^{\Lambda\, +})]\, ,
\end{equation}

\noindent
where $C^{\Lambda\, +}$ is given by Eq.~(\ref{eq:FV}). Therefore, we have the
vector field strengths written in terms of the scalars $Z^{i},X$ and the
vector $V$. Let us then consider the spacetime metric.


\subsection{The metric}
\label{sec-metric}

It is convenient to choose coordinates adapted to the timelike Killing vector
$V$ and also to use the exact 1-forms $\hat{V}^{i}$ (which, as was said before, are independent in
the timelike case) to define the spacelike coordinates. Thus, we define a time
coordinate $t$ by

\begin{equation}
V^{\mu}\partial_{\mu}\equiv\sqrt{2} \partial_{t}\, ,
\end{equation}

\noindent
and the spacelike coordinates $x^{i}$ by

\begin{equation}
\hat{V}^{i} \equiv dx^{i}\, .
\end{equation}

\noindent
Since in this case $V^{\mu}V_{\mu}=2|M|^{2}=4|X|^{2}\neq 0$, the
metric can always be constructed as

\begin{equation}
ds^{2}= |M|^{-2} [\hat{V}\otimes \hat{V} -\hat{V}^{I}{}_{J}\otimes \hat{V}^{J}{}_{I}]=
|M|^{-2} [{\textstyle\frac{1}{2}}\hat{V}\otimes \hat{V} -\hat{V}^{i}\otimes \hat{V}^{i}]\, ,
\end{equation}

\noindent
which is manifestly invariant under all the transformations leaving
invariant the equations of motion and the supersymmetry transformation
rules.  With the above choice of coordinates, the metric takes on the
form

\begin{equation}
\label{eq:metric}
ds^{2} = |M|^{2}(dt+\omega)^{2} -|M|^{-2}dx^{i}dx^{i}\, ,
\hspace{1cm}
i,j=1,2,3\, ,
\end{equation}

\noindent
where $\omega=\omega_{\underline{i}}dx^{i}$ is a time-independent 1-form that
satisfies an equation that can be found as follows: the choice of coordinates
implies

\begin{equation}
\hat{V}=2\sqrt{2}|X|^{2}(dt+\omega)\, ,  
\end{equation}

\noindent
which trivially implies that

\begin{equation}
d\omega = {\textstyle\frac{1}{2\sqrt{2}}}d(|X|^{-2}\hat{V})\, .
\end{equation}

Using Eqs.~(\ref{eq:dV}) and (\ref{eq:TV}) we find the equation for $\omega$

\begin{equation}
\label{eq:do}
d\omega = -{\textstyle\frac{i}{2\sqrt{2}}}
{}^{\star}\left[(X\mathfrak{D}X^{*}-X^{*}\mathfrak{D}X)
\wedge\frac{\hat{V}}{|X|^{4}}\right]\, .
\end{equation}

With this equation we have succeeded in expressing completely the metric in
terms of the scalars $X,Z^{i}$ and the vector $V$, as we did with the vector
field strengths.

It is convenient to define the $U(1)$ connection 1-form 

\begin{equation}
\xi\equiv {\textstyle\frac{i}{2}} \frac{X^{*}dX-XdX^{*}}{|X|^{2}}\, .  
\end{equation}

\noindent
This $\xi$ is similar to the $\xi$ defined in the $N=4,d=4$ case in
Ref.~\cite{Bellorin:2005zc}, but in this case it is exact. In terms of $\xi$
and the pullback of the K\"ahler 1-form $\mathcal{Q}$, the equation for
$\omega$ is

\begin{equation}
\label{eq:do2}
d\omega = {\textstyle\frac{1}{\sqrt{2}}}
{}^{\star}\left[(\xi-\mathcal{Q})
\wedge\frac{\hat{V}}{|M|^{2}}\right]\, .
\end{equation}

In terms of the 3-dimensional Euclidean metric, this equation takes the form 

\begin{equation}
\label{eq:do23-d}
(d\omega)_{mn} =\frac{1}{|X|^{2}}\epsilon_{mnp}
(\mathcal{Q}_{p}-\xi_{p})\, ,  
\end{equation}

\noindent
and we will later rewrite it to a more standard form.

In what follows the integrability condition for this equation will be
needed: It reads

\begin{equation}
\label{eq:dointegrability}
\partial_{m}\left[\frac{(\mathcal{Q}_{m}-\xi_{m})}{|X|^{2}}\right]=0\, .
\end{equation}


\subsection{Solving the Killing spinor equations}
\label{sec-solvingKSEtimelike}

We are now going to show that the field configurations of $N=2,d=4$ given by
the metric Eqs.~(\ref{eq:metric}) and (\ref{eq:do23-d}) and field strengths
Eqs.~(\ref{eq:decomposition2}) and (\ref{eq:FV}) are supersymmetric for
arbitrary values of the complex scalars $X,Z^{i}$ ($X\neq 0$). 

Let us start by the gauginos supersymmetry transformations
Eqs.~(\ref{eq:KSE2}). Using Eqs.~(\ref{eq:decomposition2}), we find

\begin{equation}
\not\!F^{\Lambda\, +}= -\frac{1}{|X|^{2}}C^{\Lambda\,
  +}{}_{\rho}V_{\sigma}\gamma^{\rho\sigma} 
{\textstyle\frac{1}{2}}(1-\gamma_{5})\, .  
\end{equation}

On the other hand, using the properties Eqs.~(\ref{eq:esa}) and
(\ref{eq:esaotra}), we find that 

\begin{equation}
\mathcal{T}^{i}{}_{\Lambda}C^{\Lambda\, +}{}_{\mu}= 
{\textstyle\frac{1}{2}}X^{*}\partial_{\mu}Z^{i}\, ,
\end{equation}

\noindent
and, combining this with the previous result we get

\begin{equation}
\mathcal{T}^{i}{}_{\Lambda}\not\!F^{\Lambda\, +}\epsilon^{IJ}\epsilon_{J}=
-\frac{M^{IJ}}{2|X|^{2}}\partial_{\rho}Z^{i}V_{\sigma}
\gamma^{\rho\sigma}\epsilon_{J} = 
i\not\! \partial Z^{i} (i\gamma_{0}e^{-i\alpha} \epsilon^{IJ}\epsilon_{J})\, ,
\end{equation}

\noindent
where $\alpha$ is the phase of the complex scalar bilinear $X$ and we
have used that in our Vierbein basis $\hat{V}=2|X| e_{0}$,
Eq.~(\ref{eq:VdZ}). 

Eq.~(\ref{eq:KSE2}) takes, then the form

\begin{equation}
i\not\! \partial Z^{i} 
(\epsilon^{I}+i\gamma_{0}e^{-i\alpha} \epsilon^{IJ}\epsilon_{J})=0\, ,  
\end{equation}

\noindent
and can always be solved by imposing the constraint

\begin{equation}
\label{eq:constraint}
\epsilon_{I} +i\gamma_{0}e^{i\alpha}
  \epsilon_{IJ}\epsilon^{J}=0\, ,  
\end{equation}

\noindent
which breaks half of the available supersymmetries.

Let us now consider the $0^{th}$ component of the gravitino supersymmetry
transformations Eq.~(\ref{eq:KSE1}): using Eq.~(\ref{eq:do23-d}), we find

\begin{equation}
\mathfrak{D}_{0}\epsilon_{I}= \frac{1}{\sqrt{2}|X|} 
\{\partial_{t} -X^{*}\mathfrak{D}_{m}X\gamma_{0m}\}\epsilon_{I}\, .
\end{equation}

On the other hand, using Eqs.~(\ref{eq:esta}) and (\ref{eq:esa}), we find

\begin{equation}
\mathcal{T}_{\Lambda}F^{\Lambda\, +}{}_{0m}={\textstyle\frac{i}{\sqrt{2}}} 
\mathfrak{D}_{m}X\, ,  
\end{equation}

\noindent
and combining this with the previous result we find that the $0^{th}$ component
of Eq.~(\ref{eq:KSE1}) takes, up to a global factor, the
form
\begin{equation}
\partial_{t}\epsilon_{I} 
-\frac{X^{*}\mathfrak{D}_{m}X}{|X|^{2}}\gamma_{0m}  
\left[\epsilon_{I} +i\gamma_{0}e^{i\alpha}
  \epsilon_{IJ}\epsilon^{J}\right]=0\, ,
\end{equation}

\noindent
which is always solved by time-independent spinors satisfying the
constraint (\ref{eq:constraint}).

Finally, let us consider the $m^{th}$ component of Eq.~(\ref{eq:KSE1}): using
essentially the same properties, we find on the one hand

\begin{equation}
\mathfrak{D}_{m}\epsilon_{I}= \sqrt{2}|X|
\left\{
\mathfrak{D}_{m} -{\textstyle\frac{i}{2}}\epsilon_{mnp}
-\frac{X^{*}\mathfrak{D}_{p}X}{|X|^{2}}\gamma_{0n}
\right\}\epsilon_{I}\, ,
\end{equation}

\noindent
and on the other,

\begin{equation}
\mathcal{T}_{\Lambda}F^{\Lambda\, +}{}_{ma}\gamma^{a}=
{\textstyle\frac{1}{\sqrt{2}}}(\delta_{mp}-i\epsilon_{mnp}\gamma_{0m})
\mathfrak{D}_{p}X i\gamma_{0}\, .  
\end{equation}

\noindent
Combining these two results and using the constraint Eq.~(\ref{eq:constraint}) as
to have an equation involving spinors of the same chirality, we find that the
$m^{th}$ component of Eq.~(\ref{eq:KSE1}), up to a multiplicative factor, reads

\begin{equation}
\partial_{m}(X^{-1/2}\epsilon_{I})=0\, ,  
\end{equation}

\noindent
which is solved by 

\begin{equation}
\epsilon_{I}=X^{1/2}\epsilon_{I\, 0}\, ,
\hspace{1cm}
\partial_{\mu}\epsilon_{I\, 0}=0\, ,
\hspace{1cm}  
\epsilon_{I\, 0} +i\gamma_{0}
  \epsilon_{IJ}\epsilon^{J}{}_{0}=0\, .  
\end{equation}

This is the form of the Killing spinor associated to the field
configurations that we have found. All of them are, therefore,
supersymmetric and preserve, at least, $1/2$ of the possible
supersymmetries.

Observe, however, that in this proof we have assumed that
Eq.~(\ref{eq:do23-d}) can be solved. Thus, we have assumed implicitly that the
integrability condition of this equation, Eq.~(\ref{eq:dointegrability}), has
been solved. This equation is the only condition that the field
configurations considered need to satisfy in order to be supersymmetric and,
in fact, to be well defined.  We will reconsider this condition when we
consider supersymmetric \textit{solutions}, but we can already
see that our result is different from the one in Ref.~\cite{Behrndt:1997ny},
where the pull-back of the K\"ahler form had to vanish in order for the 
solution to be supersymmetric.


\subsection{Equations of motion}
\label{sec-eoms}

Following Refs.~\cite{Tod:1995jf,Bellorin:2005zc} we are going to
introduce an $Sp(2\bar{n},\mathbb{R})$ vector of electric and magnetic
scalar potentials $E$ defined by

\begin{equation}
\nabla_{\mu}E\equiv V^{\nu}F_{\nu\mu}\, ,
\end{equation}

\noindent
where $F_{\mu\nu}$ the $Sp(2\bar{n},\mathbb{R})$ vector of field
strengths defined in Eq.~(\ref{eq:2formvector}).  Let us also define
the real symplectic sections $\mathcal{I}$ and $\mathcal{R}$

\begin{equation}
\label{eq:realsections}
\mathcal{R}\equiv \Re{\rm e}(\mathcal{V}/X)\, ,
\hspace{1.5cm}
\mathcal{I}\equiv \Im{\rm m}(\mathcal{V}/X)\, ,
\end{equation}

\noindent
where $\mathcal{V}$ is the symplectic section defined in
Appendix~\ref{sec-specialgeometry}. 

Then, using Eq.~(\ref{eq:FV}) we find

\begin{equation}
E= 2|X|^{2}\mathcal{R}\, ,
\end{equation}




%

\noindent
and using the explicit form of $\hat{V}$ and the property
Eq.~(\ref{eq:do}) we can write

\begin{equation}
\label{eq:esas}
F = -{\textstyle\frac{1}{2}} \{d[\mathcal{R} \hat{V}] 
-{}^{\star}[d\mathcal{I}\wedge \hat{V}] \}\, .
\end{equation}


\noindent
which immediately leads to the following form of the Bianchi
identities and Maxwell equations:

\begin{equation}
\label{eq:MB}
dF = {\textstyle\frac{1}{2}} d{}^{\star}[d\mathcal{I}\wedge \hat{V}]\, .
\end{equation}


Rewriting these equations in standard Cartesian $\mathbb{R}^{3}$
language, we find that the Maxwell equations and Bianchi identities
(whence, according to the KSIs, all the equations of motion of
$N=2,d=4$ supergravity) are satisfied if

\begin{equation}
\label{eq:laplaceI}
\partial_{m}\partial_{m} \mathcal{I}=0\, ,  
\end{equation}

\noindent
{\em i.e.\/} if the imaginary parts of $\mathcal{L}^{\Lambda}/X$ and
$\mathcal{M}_{\Lambda}/X$ are given by $2\bar{n}$ real harmonic
functions on $\mathbb{R}^{3}$.


Let us now reconsider the integrability condition
Eq.~(\ref{eq:dointegrability}) of the differential equation that
defines the 1-form $\omega$.  The equation for the 1-form $\omega$
(\ref{eq:do23-d}) can be rewritten as to give

\begin{equation}
\label{eq:oidi}
(d\omega)_{mn} =2\epsilon_{mnp}
\langle\,\mathcal{I}\mid \partial_{p}\mathcal{I}\, \rangle\, ,  
\end{equation}

\noindent
and its integrability condition takes on the simple form

\begin{equation}
\langle\, \mathcal{I}\mid \partial_{p}\partial_{p}\mathcal{I}\, \rangle =0\, ,
\end{equation}

\noindent
which is, as discussed in the introduction, a non-trivial condition
due to the presence of singularities in harmonic function
\cite{Denef:2000nb,Bates:2003vx}.

Summarizing, we have just shown that the configurations of $N=2,d=4$
supergravity given by the metric Eqs.~(\ref{eq:metric}) and
(\ref{eq:do23-d}) and field strengths Eqs.~(\ref{eq:decomposition2})
and (\ref{eq:FV}) are solutions of the equations of motion iff the
scalars $X,Z^{i}$ satisfy the condition Eq.~(\ref{eq:laplaceI}). The
integrability condition of the equation for the 1-form $\omega$, which
was the only condition necessary to have supersymmetry, is
automatically satisfied for supersymmetric solutions.

We still have to show how, given the real harmonic section
$\mathcal{I}$, we can express the scalars, the vector
field strengths and the metric in terms of this harmonic section: 
the metric Eq.~(\ref{eq:metric}) depends
of the 1-form $\omega$ which can be calculated from $\mathcal{I}$
integrating Eq.~(\ref{eq:oidi}) and on the absolute value of the
bilinear scalar $X$, which can be computed from $\mathcal{I}$ and
$\mathcal{R}$ observing that

\begin{equation}
\label{eq:VbarVvgl}
\langle\,(\mathcal{V}/X)^{*} \mid \mathcal{V}/X \, \rangle=
2i \langle\,\mathcal{R}\mid \mathcal{I}\, \rangle = i\frac{1}{|X|^{2}}\, ,
\end{equation}

\noindent
and that 

\begin{equation}
\langle\,(\mathcal{V}/X)^{*} \mid \mathcal{V}/X \, \rangle =
i e^{-\mathcal{K}[\mathcal{V}/X,(\mathcal{V}/X)^{*}]}\, ,
\end{equation}

\noindent
where $\mathcal{K}[\mathcal{V}/X,(\mathcal{V}/X)^{*}]$ is the
expression for the K\"ahler potential obtained in
Eq.~(\ref{eq:kpotential}) where the coordinates $\mathcal{X}$ have
been substituted by $\mathcal{L}/X$. This leads to the expression

\begin{equation}
\label{eq:Msquared}
|M|^{2}=2|X|^{2}= 2 e^{\mathcal{K}[\mathcal{V}/X,(\mathcal{V}/X)^{*}]}\, .  
\end{equation}

Observe that the form of the K\"ahler potential that has to be used
here, namely Eq.~(\ref{eq:kpotential}), is fixed after a section has
been chosen and no K\"ahler transformations are allowed. Otherwise,
the whole construction would be inconsistent since, as we have
discussed, the spacetime metric is invariant under all the symmetries
of the equations of motion and, in particular, under K\"ahler
transformations.

It is also possible to rewrite Eq.~(\ref{eq:oidi}) for $\omega$ as

\begin{equation}
(d\omega)_{mn} =\epsilon_{mnp}e^{-\mathcal{K}
[\mathcal{V}/X,(\mathcal{V}/X)^{*}]} 
\mathcal{Q}_{p}[\mathcal{V}/X,(\mathcal{V}/X)^{*}]\, .  
\end{equation}

It is clear that, in order to find $|M|^{2}$, and also in order to
find the field strengths using Eqs.~(\ref{eq:esas}) and the scalars
using, for instance, the special coordinates

\begin{equation}
Z^{i}=\frac{\mathcal{L}^{i}}{\mathcal{L}^{0}}= 
\frac{\mathcal{L}^{i}/X}{\mathcal{L}^{0}/X}\, , 
\end{equation}

\noindent
we need the real section $\mathcal{R}$ expressed as a function of
$\mathcal{I}$. Finding $\mathcal{R}(\mathcal{I})$ is equivalent to
solving the so-called ``{\sl stabilization equations}''
\cite{Ferrara:1996dd,Ferrara:1996um} which are nothing but
Eqs.~(\ref{eq:MB}) for static, spherical, asymptotically flat black
holes evaluated at the black-hole horizon: the l.h.s.~ gives the
$Sp(2\bar{n},\mathbb{R})$ vector of charges $q^{t} \equiv \left(
p^{\Lambda}\ ,\ q _{\Lambda}
\right)$ and this essentially determines the coefficient of the
harmonic functions $\mathcal{I}$. Solving the stability equations
amounts to finding the real part of the section $\mathcal{V}/X$ as a
function of the imaginary part, i.e.~of the charges.  There seems to
be no systematic procedure to find $\mathcal{R}(\mathcal{I})$ and the
solution of this problem is only known in a few simple cases, some of which we
review in Appendix~\ref{appsec:SpecShit}.

At this point we should compare our results with those of
Refs.~\cite{Behrndt:1997ny} and \cite{LopesCardoso:2000qm}.  In th
efirst of these references, the supersymmetric configurations we have
found were proposed as an \textit{Ansatz} and they were shown to be
supersymmetric. In the second, the same solutions were found from the
KSEs of superconformal gravity starting with an \textit{Ansatz} for
the contraint satisfied by the Killing spinors of the form
Eq.~(\ref{eq:constraint})\footnote{Their results include also $R^{2}$
  corrections, but we are not concerned with them here.}.  We have
just shown that all the solutions\footnote{The exceptions are the
  maximally supersymmetric Minkowski and Bertotti-Robinson-type
  solutions.} in the timelike class satisfy this constraint and there
are no more solutions than those found by Behrndt, L\"ust and Sabra in
the timelike class, although there are more supersymmetric solutions
in the null class, as we are going to see.


\section{The null case}
\label{sec-null}

In the null case\footnote{The technical details concerning the
  normalization of the spinors and the construction of the bilinears
  in this case are explained in the Appendix of
  Ref.~\cite{Bellorin:2005zc}.} the two spinors $\epsilon_{I}$ are
proportional $\epsilon_{I}=\phi_{I}\epsilon$. The complex scalar
functions $\phi_{I}$ carry a -1 $U(1)$ charge w.r.t.~the purely imaginary
connection

\begin{equation}
\zeta\equiv \phi^{I} d\phi_{I}\, ,
\end{equation}

\noindent
opposite to that of the spinor $\epsilon$, so the $\epsilon_{I}$ are
neutral. On the other hand, the $\phi_{I}$s are neutral with respect
to the K\"ahler connection, and the K\"ahler weight of the spinor
$\epsilon$ is the same as that of the spinor $\epsilon_{I}$,
i.e.~$1/2$.

We are now going to substitute $\epsilon_{I}=\phi_{I}\epsilon$ into
the KSEs and we are going to use the normalization condition of the
scalars $\phi_{I}\phi^{I}=1$ to split the KSEs into three algebraic
and one differential equation for $\epsilon$; one of the algebraic
equations for $\epsilon$ will be a differential equation for
$\phi_{I}$.

This substitution immediately yields

\begin{eqnarray}
\partial_{\mu} \phi_{I} \epsilon+\phi_{I} \mathfrak{D}_{\mu}\epsilon 
-\epsilon_{IJ}\phi^{J} T^{+}{}_{\mu\nu}
\gamma^{\nu} \epsilon^{*} & = & 0\, , 
\label{eq:gravitinodegenerate} \\
& & \nonumber \\
\phi^{I}\!\not\!\partial Z^{i} \epsilon^{*} +\epsilon^{IJ}\phi_{J}
\not\! G^{i\, +} \epsilon & = & 0 \, . 
\label{eq:dilatinodegenerate}
\end{eqnarray}

Acting on Eq.~(\ref{eq:gravitinodegenerate}) with $\phi^{I}$ leads to

\begin{equation}
\mathfrak{D}_{\mu}\epsilon = -\phi^{I} \partial_{\mu} \phi_{I} \epsilon\, ,
\end{equation}

\noindent
which takes the form

\begin{equation}
\label{eq:Depsilon}
\tilde{\mathfrak{D}}_{\mu}\epsilon\equiv 
(\mathfrak{D}_{\mu} +\zeta_{\mu})\epsilon=0\, ,  
\end{equation}

\noindent
and becomes the only differential equation for $\epsilon$. Observe
that the covariant derivative $\tilde{\cal D}_{\mu}$ contains, apart
from the connection $\zeta$, the spin and K\"ahler connections. Plugging
Eq.~(\ref{eq:Depsilon}) into Eq.~(\ref{eq:gravitinodegenerate}) as to
eliminate $\mathcal{D}_{\mu}\epsilon$ we obtain

\begin{equation}
\label{eq:gravitinodegII}
\tilde{\mathfrak{D}}\phi_{I}
\epsilon  +\epsilon_{IJ}\phi^{J} T^{+}{}_{\mu\nu}\gamma^{\nu} 
\epsilon^{*}= 0\, ,
\hspace{2cm}
(\tilde{\mathfrak{D}}\phi_{I} \equiv (\partial_{\mu} -\zeta_{\mu})\phi_{I})\, ,
\end{equation}

\noindent
which is one of the algebraic constraints for $\epsilon$ and is a differential
equation for $\phi_{I}$.

Multiplying Eq.~(\ref{eq:dilatinodegenerate}) with $\phi_{I}$, we see that it
splits into two algebraic constraints for $\epsilon$:

\begin{eqnarray}
\not\!\partial Z^{i} \epsilon^{*} & = & 0\, , \label{eq:dtaueps} \\
& & \nonumber \\
\not\! G^{i\, +}\epsilon & = & 0\, . \label{eq:vinF-}
\end{eqnarray}

Finally, we add to the system an auxiliary spinor $\eta$, with the
same chirality as $\epsilon$ but with all $U(1)$ charges opposite to
those of $\epsilon$ and normalized by the condition

\begin{equation}
\bar{\epsilon}\eta={\textstyle\frac{1}{2}}\, .  
\end{equation}

\noindent
This normalization condition will be preserved if and only if $\eta$
satisfies

\begin{equation}
\label{eq:Deta}
\tilde{\mathfrak{D}}_{\mu}\eta +a_{\mu}\epsilon=0\, ,  
\end{equation}

\noindent
for some $a_{\mu}$ with $U(1)$ charges $-2$ times those of $\epsilon$, {\em i.e.}

\begin{equation}
\tilde{\mathfrak{D}}_{\mu}a_{\nu} =
(\nabla_{\mu} -2\zeta_{\mu} -i\mathcal{Q}_{\mu})a_{\nu}\, ,
\end{equation}

\noindent
to be determined by the requirement that the integrability conditions
of this differential equation have to be compatible with those of the
differential equation for $\epsilon$.

Observe that the null tetrad of vector bilinears one constructs
from $\epsilon$ and $\eta$ will in general have non-trivial charges
and, in particular, non-trivial K\"ahler weight: taking into account
the definition of the bilinear vectors in Ref.~\cite{Bellorin:2005zc},
which we reproduce here for convenience

\begin{equation}
\label{eq:nulltetraddef}
l_{\mu}=i\sqrt{2}\bar{\epsilon^{*}}\gamma_{\mu}\epsilon\, ,
\hspace{.5cm}
n_{\mu}=i\sqrt{2}\bar{\eta^{*}}\gamma_{\mu}\eta\, ,
\hspace{.5cm}
m_{\mu}=i\sqrt{2}\bar{\epsilon^{*}}\gamma_{\mu}\eta=
i\bar{\eta}\gamma_{\mu}\epsilon^{*}\, ,
\hspace{.5cm}
m_{\mu}^{*}=i\sqrt{2}\bar{\epsilon}\gamma_{\mu}\eta^{*}=
i\bar{\eta^{*}}\gamma_{\mu}\epsilon\, .
\end{equation}

\noindent
we see that $l$ and $n$ have 0 $U(1)$ charges but $m$ has $-2$ times
the charges of $\epsilon$ and $m^{*}$ has $+2$ times the charges of
$\epsilon$. The  metric

\begin{equation}
\label{eq:nullcasemetric}
ds^{2}= 2\hat{l}\otimes \hat{n} -2\hat{m}\otimes \hat{m}^{*}\, ,  
\end{equation}

\noindent
is invariant, though.

The orientation of the null tetrad is important: we choose the
relation between a standard Cartesian tetrad
$\{e^{0},e^{1},e^{2},e^{3}\}$ and the complex null tetrad
$\{e^{u},e^{v},e^{z},e^{z^{*}}\}=\{\hat{l},\hat{n},\hat{m},\hat{m}^{*}\}$
to be

\begin{equation}
\left(
  \begin{array}{c}
e^{u} \\
e^{v} \\
e^{z} \\
e^{z^{*}} \\
  \end{array}
\right) 
=
\frac{1}{\sqrt{2}}
\left(
  \begin{array}{rr|rr}
1 & 1 & & \\
1 & -1 & & \\
\hline
& & 1 & i \\
& & 1 & -i \\
  \end{array}
\right) 
\left(
  \begin{array}{c}
e^{0} \\
e^{1} \\
e^{2} \\
e^{3} \\
  \end{array}
\right)\, . 
\end{equation}

This translates into identical relations between gamma matrices:

\begin{equation}
\left(
  \begin{array}{c}
\gamma^{u} \\
\gamma^{v} \\
\gamma^{z} \\
\gamma^{z^{*}} \\
  \end{array}
\right) 
=
\left(
  \begin{array}{c}
\not l \\
\not n \\
\not\! m \\
\not\! m^{*} \\
  \end{array}
\right) 
=
\frac{1}{\sqrt{2}}
\left(
  \begin{array}{rr|rr}
1 & 1 & & \\
1 & -1 & & \\
\hline
& & 1 & i \\
& & 1 & -i \\
  \end{array}
\right) 
\left(
  \begin{array}{c}
\gamma^{0} \\
\gamma^{1} \\
\gamma^{2} \\
\gamma^{3} \\
  \end{array}
\right)\, . 
\end{equation}

This choice implies for the chirality matrix 

\begin{equation}
\gamma_{5}\equiv -i \gamma^{0}\gamma^{1}\gamma^{2}\gamma^{3} =
-\gamma^{uv}\gamma^{zz^{*}}\, .  
\end{equation}


\subsection{Killing equations for the vector bilinears and first consequences}

We are now ready to derive equations involving the bilinears, in
particular the vector bilinears constructed from $\epsilon$ and
the auxiliary spinor $\eta$ introduced above. First we deal with the
equations that do not involve derivative of the spinors. Acting with
$\bar{\epsilon}$ on Eq.~(\ref{eq:gravitinodegII}) and with
$\bar{\epsilon}\gamma^{\mu}$ on Eq.~(\ref{eq:vinF-}) we get, respectively

\begin{eqnarray}
T^{+}{}_{\mu\nu}l^{\nu} & = & 0\, ,\\
& & \nonumber \\
G^{i\, +}{}_{\mu\nu}l^{\nu} & = & 0\, ,
\end{eqnarray}

\noindent
which together imply

\begin{equation}
F^{\Lambda\, +} {}_{\mu\nu}l^{\nu}  = 0\, , 
\end{equation}

\noindent
which in its turn implies

\begin{equation}
\label{eq:FL}
F^{\Lambda\, +}={\textstyle\frac{1}{2}}\phi^{\Lambda} 
\hat{l}\wedge \hat{m}^{*}\, ,   
\end{equation}

\noindent
where $\phi^{\Lambda}$ is some complex function. This form of
$F^{\Lambda\, +}$ completely solves Eq.~(\ref{eq:vinF-}), as becomes
paramount through the Fierz identity

\begin{equation}
l_{\mu}\gamma^{\mu\nu}\epsilon^{*}=3l^{\nu}\epsilon^{*}\, .
\end{equation}

Acting with $\bar{\eta}$ on Eq.~(\ref{eq:gravitinodegII}) we get

\begin{equation}
\label{eq:gravitinodegIII}
\tilde{\mathfrak{D}}_{\mu}\phi_{I}
+i\sqrt{2}\epsilon_{IJ}\phi^{J} T^{+}{}_{\mu\nu}m^{\nu}= 0\, ,
\end{equation}

\noindent
and substituting Eq.~(\ref{eq:FL}) into  it, we obtain

\begin{equation}
\label{eq:gravitinodegIV}
\tilde{\mathfrak{D}}_{\mu}\phi_{I}
-{\textstyle\frac{i}{\sqrt{2}}}\epsilon_{IJ}\phi^{J}\mathcal{T}_{\Lambda} 
\phi^{\Lambda}l_{\mu}= 0\, .
\end{equation}

Finally, acting with $\bar{\epsilon}$ and $\bar{\eta}$ on
Eq.~(\ref{eq:dtaueps}) we get

\begin{eqnarray}
l^{\mu}\partial_{\mu}Z^{i} & = & 0\, , \label{eq:ldt}\\
& & \nonumber \\
m^{\mu}\partial_{\mu}Z^{i} & = & 0\, , \label{ldm}
\end{eqnarray}

\noindent
which imply

\begin{equation}
\label{eq:dZ}
dZ^{i}=A^{i}\hat{l}+B^{i}\hat{m}\, ,  
\end{equation}

\noindent
for some functions $A^{i}$ and $B^{i}$ that are $v$ independent.
Observe that, since $dZ^{i}$ and $\hat{l}$ have no K\"ahler weight and
$\hat{m}$ has K\"ahler weight $+2$, $B^{i}$ must have K\"ahler weight
$-2$. As shown in Refs.~\cite{Tod:1995jf,Bellorin:2005zc}, for a
single scalar ($dZ=A\hat{l}+B\hat{m}$) we can always assume that either
$B$ is zero (case $A$) or $A$ is zero (case $B$). However, for more
than one scalar, it is not possible to remove all the $A^{i}$s and we
are going to have, in general, non-vanishing $A^{i}$s and $B^{i}$s,
although we can consider particular cases in which either all the
$A^{i}$s or all the $B^{i}$s vanish.

Observe that, due to the Fierz identity

\begin{equation}
\label{eq:mlFierz}
\not l\epsilon^{*}=\not\!m\epsilon^{*}=0\, ,  
\end{equation}

\noindent
the above expression solves Eq.~(\ref{eq:dtaueps}) identically.
These are all the algebraic equations for the bilinears.  Now, from
Eqs.~(\ref{eq:Depsilon}) and (\ref{eq:Deta}) we find the differential
equations

\begin{eqnarray}
\label{eq:dtetrad1}
  \nabla_{\mu} l_{\nu} & = & 0\, ,\\
& & \nonumber \\
\label{eq:dtetrad2}
  \tilde{\mathfrak{D}}_{\mu} n_{\nu} & = & \nabla_{\mu}n_{\nu}=
-a^{*}_{\mu}m_{\nu} -a_{\mu}m^{*}_{\nu}\, ,\\
& & \nonumber \\
\label{eq:dtetrad3}
  \tilde{\mathfrak{D}}_{\mu} m_{\nu} & = & 
(\nabla_{\mu}-2\zeta_{\mu} -i\mathcal{Q}_{\mu})m_{\nu} 
=-a_{\mu}l_{\nu}\, .
\end{eqnarray}


\subsection{Equations of motion and integrability constraints}
\label{eq:eomintegrabilityconstraintsnull}

Our immediate objective is to find information about the connection
$\zeta_{\mu}$ using the KSIs and the integrability equations of
Eqs.~(\ref{eq:Depsilon}) and (\ref{eq:Deta}).

Using the results of the previous section, we can write the Einstein
equations the form

\begin{equation}
  \begin{array}{rcl}
\mathcal{E}_{\mu\nu}
-{\textstyle\frac{1}{2}}g_{\mu\nu}\mathcal{E}^{\rho}{}_{\rho}
& = &
R_{\mu\nu} 
+\left[
2\mathcal{G}_{ij^{*}}A^{i}A^{*\, j^{*}}
-8\Im{\rm m}\mathcal{N}_{\Lambda\Sigma} \phi^{\Lambda}\phi^{*\, \Sigma}
\right]
l_{\mu}l_{\nu}\\
& & \\
& & 
+2\mathcal{G}_{ij^{*}}B^{i}B^{*\, j^{*}}
m_{(\mu}m^{*}_{\nu)}
+2\mathcal{G}_{ij^{*}}A^{i}B^{*\, j^{*}}l_{(\mu}m^{*}_{\nu)}
\\
& & \\
& & 
+2\mathcal{G}_{ij^{*}}B^{i}A^{*\, j^{*}}l_{(\mu}m_{\nu)}
\, .
\end{array}
\end{equation}

Comparing with the KSI Eq.~(\ref{eq:ksinull1}), we end up with tho following two conditions

\begin{eqnarray}
\label{eq:inte1}
R_{\mu\nu}l^{\nu} & = & 0\, ,\\
& & \nonumber \\
\label{eq:inte2}
R_{\mu\nu}m^{\nu} - \mathcal{G}_{ij^{*}}(A^{i}l_{\mu}+B^{i}m_{\mu})
B^{*\, j^{*}} & = & 0\, .
\end{eqnarray}

Commuting the derivative and projecting with gamma matrices and
spinors in the usual way, and using

\begin{equation}
(d\mathcal{Q})_{\mu\nu} m^{* \nu} 
=i\mathcal{G}_{ij^{*}}B^{i}B^{*\, j^{*}}m^{*}_{\mu}\, ,
\hspace{1cm}  
(d\mathcal{Q})_{\mu\nu}l^{\nu}=(d\mathcal{Q})_{\mu\nu}n^{\nu}=0\, ,
\end{equation}

\noindent
which follow from the definition of the K\"ahler connection and K\"ahler form
Eq.~(\ref{eq:dZ}), it is easy to find, from Eq.~(\ref{eq:Depsilon})

\begin{eqnarray}
\{R_{\mu\nu} +2 (d\zeta)_{\mu\nu}\} l^{\nu} & = & 0\, ,\\
& & \nonumber \\
\{R_{\mu\nu} +2 (d\zeta)_{\mu\nu}\} m^{*\, \nu}
-\mathcal{G}_{ij^{*}}B^{i}(A^{*\, j^{*}}l_{\mu}
+B^{*\, j^{*}}m^{*}_{\mu}) & = & 0\, ,
\end{eqnarray}

\noindent
and from Eq.~(\ref{eq:Deta})

\begin{eqnarray}
\{R_{\mu\nu} -2 (d\zeta)_{\mu\nu}\}m^{\nu} 
-\mathcal{G}_{ij^{*}}(A^{i}l_{\mu}+B^{i}m_{\mu})B^{*\, j^{*}}
+2(\tilde{\mathfrak{D}}a)_{\mu\nu}l^{\nu} & = & 0\, ,\\
& & \nonumber \\
\{R_{\mu\nu} -2 (d\zeta)_{\mu\nu}\}n^{\nu} 
+2(\tilde{\mathfrak{D}}a)_{\mu\nu}m^{*\, \nu} & = & 0\, .
\end{eqnarray}

Comparing these three sets of equations, we find that they are
compatible if

\begin{equation}
(d\zeta)_{\mu\nu}l^{\nu} = (d\zeta)_{\mu\nu}m^{\nu} =0\, ,\,\,\,\,
\Rightarrow
d\zeta=0\, ,\,\,\,\,
\Rightarrow \zeta = d\alpha\, ,
\end{equation}

\noindent
locally, and, eliminating $\zeta$ by a local phase redefinition,
$\tilde{\mathfrak{D}}a$ becomes just $\mathfrak{D}a$ and we get

\begin{eqnarray}
(\mathfrak{D}\hat{a})_{\mu\nu} l^{\nu} & = &  0\, ,\\
& & \nonumber \\
(\mathfrak{D}\hat{a})_{\mu\nu} m^{*\, \nu} & = &  
-{\textstyle\frac{1}{2}}R_{\mu\nu}n^{\nu}\, ,
\end{eqnarray}

\noindent
so that 

\begin{equation}
\label{eq:da}
\mathfrak{D}\hat{a}= -{\textstyle\frac{1}{2}} R_{z^{*}u}
\hat{m}\wedge\hat{m}^{*}
+{\textstyle\frac{1}{2}} R_{uu}\hat{l}\wedge \hat{m}
+C\hat{l}\wedge \hat{m}^{*}\, ,
\end{equation}

\noindent
where $C$ is a function that needs to be chosen as to make this equation (and,
hence, Eq.~(\ref{eq:Deta})) integrable. We also have to satisfy the
integrability equations (\ref{eq:inte1}) and (\ref{eq:inte2}).

Another consequence of the elimination of $\zeta_{\mu}$ is that
$\tilde{\mathfrak{D}}\phi_{I}$ becomes just $d\phi_{I}$, whence
Eq.~(\ref{eq:gravitinodegIV}) implies that $d\phi_{I}\sim \hat{l}$
and the graviphoton combination

\begin{equation}
\phi\equiv \mathcal{T}_{\Lambda} \phi^{\Lambda}\, ,
\hspace{1cm}
d\phi \sim \hat{l}\, . 
\end{equation}

Observe that a similar statement cannot be made about the matter combinations

\begin{equation}
\psi^{i}\equiv \mathcal{T}^{i}{}_{\Lambda}\phi^{\Lambda}\, .  
\end{equation}

The variables $\phi,\psi^{i}$ will be convenient for further
calculations, and the relation between them and the
$\phi^{\Lambda}$ can be obtained from Eq.~(\ref{eq:completeness}):

\begin{equation}
  \label  {eq:inverse}
\phi^{\Lambda}=  i\mathcal{L}^{*\, \Lambda}\phi
+2f^{\Lambda}{}_{i}\psi^{i}\, .
\end{equation}

Using these variables, the symplectic vector of field strengths
defined in Eq.~(\ref{eq:2formvector}) takes the form


\begin{equation}
F = \left(\mathcal{U}_{i}\psi^{i}
+\textstyle{\frac{i}{2}}\mathcal{V}^{*}\phi \right)\hat{l}\wedge \hat{m}^{*} 
+\textrm{c.c.}\, ,  
\end{equation}

\noindent
and the symplectic vector containing the Bianchi identities and
Maxwell equations, defined in Eq.~(\ref{eq:maxwellvector}) is, in
differential-form language

\begin{equation}
{}^{\star}\hat{\mathcal{E}}=dF= -\hat{l}\wedge \left[d \left(\mathcal{U}_{i}\psi^{i}
+\textstyle{\frac{i}{2}}\mathcal{V}^{*}\phi \right)\wedge \hat{m}^{*}
+\left(\mathcal{U}_{i}\psi^{i}
+\textstyle{\frac{i}{2}}\mathcal{V}^{*}\phi \right)d\hat{m}^{*}
+{\rm c.c.}
\right]\, .  
\end{equation}

\noindent
Since $d\phi\sim \hat{l}$, it drops out of the above equations.
Next, we substitute

\begin{equation}
d \mathcal{V}^{*} = \mathcal{U}^{*}{}_{i^{*}}dZ^{*\, i^{*}} 
+\textstyle{\frac{1}{2}}\mathcal{V}^{*}d\mathcal{K}\, .
\end{equation}

\noindent
Finally, using Eqs.~(\ref{eq:dtetrad3}) and (\ref{eq:dZ}) we find

\begin{equation}
\hat{l}\wedge d\hat{m}^{*}= \hat{l}\wedge(-{\textstyle\frac{1}{2}}d\mathcal{K})
\wedge \hat{m}^{*}\, ,
\end{equation}

\noindent
which, after substituting and assuming independence of $v$, leads to

\begin{equation}
\label{eq:MBnull}
{}^{\star}\hat{\mathcal{E}}=e^{\mathcal{K}/2}
d(e^{-\mathcal{K}/2}\psi^{i}\mathcal{U}_{i} )\wedge
\hat{l} \wedge \hat{m}^{*}+{\rm c.c.}  
\end{equation}

\noindent
We are now in a position to check the KSIs that involve the Maxwell equations and
Bianchi identities: first of all, Eqs.~(\ref{eq:ksinull4}) are
satisfied automatically, and Eq.~(\ref{eq:ksinull3}) can be put in the
form

\begin{equation}
\langle\, {}^{\star}\hat{\mathcal{E}}  
 \mid \mathcal{V}\, \rangle  =0\, .
\end{equation}

\noindent
Rewriting 

\begin{equation}
{}^{\star}\hat{\mathcal{E}}=[e^{\mathcal{K}}d(e^{-\mathcal{K}}d\psi^{i})\mathcal{U}_{i}
-e^{\mathcal{K}/2}\psi^{i}m^{*\, \mu}\partial_{\mu}Z^{j} 
\mathfrak{D}_{j}\mathcal{U}_{i}\, \hat{m} ]\wedge
\hat{l} \wedge \hat{m}^{*}+{\rm c.c.}  
\end{equation}

\noindent 
and using $\langle \mathcal{U}_{i}\mid \mathcal{V}\rangle=
\langle \mathcal{U}^{*}{}_{i^{*}}\mid \mathcal{V}\rangle= \langle
\mathfrak{D}_{j}\mathcal{U}_{i}\mid \mathcal{V}\rangle= \langle
\mathfrak{D}_{j^{*}}\mathcal{U}^{*}{}_{i^{*}}\mid \mathcal{V}\rangle =
0$ we see  that the above equation is always satisfied.

The only component of these equations is, then, 

\begin{equation}
\label{eq:MBnull2}
m^{*\, \mu}\partial_{\mu}(e^{-\mathcal{K}/2}\psi^{i}\mathcal{U}_{i})- {\rm c.c.}=0\, . 
\end{equation}

Finally, let us consider the scalar equation of motion, which takes the
form

\begin{equation}
\label{eq:scalareom}
\mathcal{E}^{i^{*}} 
= 
m^{*\, \mu}\mathfrak{D}_{\mu}B^{*\, i^{*}} 
-B^{*\, i^{*}}
l^{\mu}a^{*}_{\mu}\, .
\end{equation}

According to Eq.~(\ref{eq:ksinull5}), this combination has to vanish
in order to have supersymmetry, and in the next section we are going
to see that this happens if the $B^{i}$s are covariantly holomorphic
in a complex coordinate, denoted by $z$, and $l^{\mu}a_{\mu}=0$.


\subsection{Metric}
\label{sec:NullMetric}
In order to advance and check the KSIs involving the Ricci tensor we need an
explicit form of the metric. This form is dictated by the existence of
a covariantly constant null Killing vector,
Eq.~(\ref{eq:dtetrad1}), which tells us that the spacetime is a
Brinkmann $pp$-wave, \cite{kn:Br1,kn:Br2}.  Since $l^{\mu}$ is a
Killing vector and $d\hat{l}=0$ we can introduce the coordinates $u$
and $v$ such that

\begin{eqnarray}
\hat{l}= l_{\mu}dx^{\mu} & \equiv & du\, , \label{u}\\
& & \nonumber \\
l^{\mu}\partial_{\mu}  & \equiv & \frac{\partial}{\partial v}\, . \label{v}
\end{eqnarray}

\noindent
We can also define a complex coordinate $z$ by 

\begin{equation}
\label{eq:z}
\hat{m} = e^{U}dz\, ,  
\end{equation}

\noindent
where $U$ may depend on $z,z^{*}$ and $u$. Eq.~(\ref{eq:ldt}) then states
that the scalars $Z^{i}$ are functions of $z$ and $u$ only:

\begin{equation}
Z^{i}=Z^{i}(z,u)\, ,
\end{equation}

\noindent
and, therefore, the functions $A^{i}$ and $B^{i}$ defined in
Eq.~(\ref{eq:dZ}) are

\begin{equation}
\label{eq:functionsAB}
A^{i}=\partial_{\underline{u}}Z^{i}\, ,
\hspace{1cm}
e^{U}B^{i}= \partial_{\underline{z}}Z^{i}\, ,\,\,\, \Rightarrow 
\partial_{\underline{z}^{*}} (e^{U}B^{i})=0\, .
\end{equation}

Finally, the most general form that $\hat{n}$ can take in this case is

\begin{equation}
\hat{n}= dv + H du +\hat{\omega}\, ,   
\hspace{1cm}
\hat{\omega}=\omega_{\underline{z}}dz +\omega_{\underline{z}^{*}}dz^{*}\, ,
\end{equation}

\noindent
where all the functions in the metric are independent of $v$ and where
either $H$ or the 1-form $\hat{\omega}$ could, in principle, be removed by a
coordinate transformation but we have to check that the tetrad
integrability equations (\ref{eq:dtetrad1})-(\ref{eq:dtetrad3}) are
satisfied by our choices of $e^{U},H$ and $\hat{\omega}$.
Eq.~(\ref{eq:nullcasemetric}) and the above choice of coordinates,
lead to the metric\footnote{The components of the
  connection and the Ricci tensor of this metric can be found in the
  Appendix of Ref.~\cite{Bellorin:2005zc}.}

\begin{equation}
\label{eq:Brinkmetric}
ds^{2} = 2 du (dv + H du +\hat{\omega})
-2e^{2U}dzdz^{*}\, .
\end{equation}

Let us then consider the tetrad integrability equations
(\ref{eq:dtetrad1})-(\ref{eq:dtetrad3}): the first equation is solved
because the metric does not depend on $v$. The third equation, with
the choice of coordinate $z$, Eq.~(\ref{eq:z}),
implies\footnote{Actually, the most general solution is $U =
  -\mathcal{K}/2+h(u)$, but we can always eliminate $h(u)$ by a
  redefinition of $z$ that does not change the structure of the
  metric.}

\begin{eqnarray}
\hat{a} & = & \left[\dot{U} 
-i\mathcal{Q}_{\underline{u}}\right]\hat{m}  +D\hat{l}\, ,\\
& & \nonumber \\
\label{eq:UK}
U & = & -\mathcal{K}/2\, ,
\end{eqnarray}

\noindent
where $D(z,z^{*},u)$ is a functions to be determined and over-dots
denote partial derivation w.r.t.~$u$.  Combining both equations we
get

\begin{equation}
\hat{a} =  -A^{i}\partial_{i}\mathcal{K}\hat{m}  +D\hat{l}\, .
\end{equation}

Finally, the second tetrad integrability equation (\ref{eq:dtetrad2})
implies

\begin{eqnarray}
D & = & e^{\mathcal{K}/2}
(\partial_{\underline{z}^{*}}H -\dot{\omega}_{\underline{z}^{*}})\, ,\\
& & \nonumber \\
(d\hat{\omega})_{\underline{z}\underline{z}^{*}}
 & = & 
2i e^{-\mathcal{K}}\mathcal{Q}_{\underline{u}}\, ,
\label{eq:doQ}
\end{eqnarray}

\noindent
whence $\hat{a}$ is given by

\begin{equation}
\label{eq:exprehata}
\hat{a} =  -A^{i}\partial_{i}\mathcal{K}\hat{m}  
+e^{\mathcal{K}/2}(\partial_{\underline{z}^{*}}H 
-\dot{\omega}_{\underline{z}^{*}})\hat{l}\, . 
\end{equation}

Observe that this implies $a_{\mu}l^{\mu}=0$. On the other hand,
the last of Eqs.~(\ref{eq:functionsAB}) together with
Eq.~(\ref{eq:UK})

\begin{equation}
\partial_{\underline{z}^{*}}(e^{-\mathcal{K}/2}B^{i}) =
\mathfrak{D}_{\underline{z}^{*}}B^{i}= 0\, .
\end{equation}

\noindent
Thus, the scalar equation of motion  (\ref{eq:scalareom}) is identically satisfied
and so is the KSI~(\ref{eq:ksinull3}).

Having a coordinate system, we can check the integrability conditions
Eqs.~(\ref{eq:inte1},\ref{eq:inte2}). The first of these is
automatically satisfied for Brinkmann metrics. The second splits into

\begin{equation}
\label{eq:inte3}
  \begin{array}{rcl}
R_{uz^{*}} + \mathcal{G}_{ij^{*}}A^{i}B^{*\, j^{*}}  & = & 0\, ,\\
& & \\
R_{zz^{*}} + \mathcal{G}_{ij^{*}}B^{i}B^{*\, j^{*}}  & = & 0\, .\\
\end{array}
\end{equation}

\noindent
The coefficients of the Ricci tensor for Brinkmann metrics were given
in the Appendix of Ref.~\cite{Bellorin:2005zc}: substituting
Eqs.~(\ref{eq:UK}) and (\ref{eq:doQ}) into those expressions and using the holomorphicity of
the $Z^{i}$s the above equations are seen to be satisfied identically.\footnote{
 We would like to point out a typo in Tod's article \cite{Tod:1995jf}: The metric has the part $2m_{(\mu}m^{*}_{\nu )}$,
which together with \cite[(A.9--10)]{Tod:1995jf}, indicates that the metric factor should be $e^{-2\phi}\omega\bar{\omega}$
and not Eq. \cite[(A.17)]{Tod:1995jf}, which is just the inverse. After taking this into account, Tod's
results fully agree with the ones presented here.
}

Having an expression for $\hat{a}$, Eq.~(\ref{eq:exprehata}), we can impose the integrability
condition Eq.~(\ref{eq:da}), resulting in

\begin{eqnarray}
C & = & -e^{\mathcal{K}/2}\partial_{\underline{z}^{*}}
[e^{\mathcal{K}/2}(\partial_{\underline{z}^{*}}H 
-\dot{\omega}_{\underline{z}^{*}})]\, ,\\
& & \nonumber \\
R_{uu} & = & -2\partial_{\underline{u}}(A^{i}\partial_{i}\mathcal{K}) 
-2e^{\mathcal{K}/2}\mathfrak{D}_{\underline{z}}
[e^{\mathcal{K}/2}(\partial_{\underline{z}^{*}}H 
-\dot{\omega}_{\underline{z}^{*}})] +2(A^{i}\partial_{i}\mathcal{K})^{2} \, ,\\ 
& & \nonumber \\
R_{uz^{*}} & = & -2e^{\mathcal{K}/2}\partial_{\underline{z}^{*}}
(A^{i}\partial_{i}\mathcal{K})=-2\mathcal{G}_{ij^{*}}A^{i}B^{*\, j^{*}}\, .
\end{eqnarray}

\noindent
The second equation is satisfied automatically.  The last equation is,
however, incompatible with the integrability equation
Eq.~(\ref{eq:inte3}) and with the actual value of $R_{uz^{*}}$ for the
Brinkmann metric unless

\begin{equation}
\label{eq:extracondition1}
\partial_{\underline{u}}Z^{i}\partial_{\underline{z}^{*}}Z^{*\, j^{*}}  
\mathcal{G}_{ij^{*}}=0\, .
\end{equation}

\noindent
This conditions is a consequence of the choice of $\eta$, {\em i.e.\/} of our
frame and coordinate choice, and should be of no importance whatsoever to the
problem of solving the KSEs.  Actually, it is easy to see that it can
always be satisfied by a shift in $\eta$ that preserves the
normalization condition $\bar{\epsilon}\eta=1/2$:

\begin{equation}
\eta^{\prime}=\eta +\delta \epsilon\, ,\,\,\, \Rightarrow\,\,\,
\left\{
  \begin{array}{rcl}
\hat{l}^{\prime} & = & \hat{l}\, ,\\
\hat{n}^{\prime} & = & \hat{n}+\delta^{*}\hat{m}+ \delta \hat{m}^{*} 
+|\delta|^{2}\hat{l}\, ,\\
\hat{m}^{\prime} & = & \hat{m} +\delta \hat{l}\, .\\
\end{array}
\right.
\end{equation}

\noindent 
If $B^{i}=\partial_{z}Z^{i}=0$ then the condition is automatically
satisfied.  If $B^{i}=\partial_{z}Z^{i}\neq 0$ then
$\mathcal{G}_{ij^{*}}B^{i}B^{*\, j^{*}}\neq 0$ and we just have to
perform the above shift with

\begin{equation}
\delta = 
-\frac{\mathcal{G}_{ij^{*}}A^{i}B^{*\, j^{*}}}
{\mathcal{G}_{ij^{*}}B^{i}B^{*\, j^{*}}}\, ,
\end{equation}

\noindent
in order to trivialize the condition (\ref{eq:extracondition1}).


\subsection{Solving the Killing spinor equations}
\label{sec:NullKSE}
We are now going to see that field configurations given by a metric of
the form (Eqs.~(\ref{eq:Brinkmetric}) and (\ref{eq:UK}))

\begin{equation}
\label{eq:metricnullcase}
ds^{2} = 2 du (dv + H du +\hat{\omega})
-2e^{-\mathcal{K}}dzdz^{*}\, ,
\end{equation}

\noindent 
where $\hat{\omega}$ satisfies (Eq.~(\ref{eq:doQ})) 

\begin{equation}
\label{eq:DetOmegaQ}
(d\omega)_{\underline{z}\underline{z}^{*}}=
2i e^{-\mathcal{K}}\mathcal{Q}_{\underline{u}}\, ,
\end{equation}

\noindent
scalars of the form (Eq.~(\ref{eq:dZ}))

\begin{equation}
dZ^{i}=A^{i}\hat{l}+B^{i}\hat{m}\, ,  
\end{equation}

\noindent
and vector field strengths of the form (Eq.~(\ref{eq:FL}))

\begin{equation}
F^{\Lambda\, +}={\textstyle\frac{1}{2}}\phi^{\Lambda} 
\hat{l}\wedge \hat{m}^{*}\, ,   
\end{equation}

\noindent
are always supersymmetric, even though we derived these equations as
necessary conditions for supersymmetry.

With the above form of the scalars and vector field strengths the KSE
$\delta_{\epsilon}\lambda^{iI}=0$ takes the form

\begin{equation}
iA^{i}\not l \epsilon^{I}   +iB^{i}\not\! m \epsilon^{I} 
-{\textstyle\frac{1}{2}}\epsilon^{IJ}\mathcal{T}^{i}{}_{\Lambda}\phi^{\Lambda}
\not\! m^{*}\!\! \not l \epsilon_{J}=0\, ,   
\end{equation}

\noindent
and can be solved by imposing two conditions on the spinors:

\begin{equation}
\label{eq:nullconditions}
\not l \epsilon^{I} = 0\, ,
\hspace{1cm}
\not\! m \epsilon^{I} = 0\, ,
\end{equation}

\noindent
which formally coincide with the Fierz identities
Eqs.~(\ref{eq:mlFierz}), although now, since there is no \textit{à
priori} relation between $l$, $m$ and $\epsilon^{I}$, they are not
identities but \textit{constraints on} $\epsilon^{I}$. This fact should be
enough to show that they are compatible, but we are going to go
further and show that they are equivalent. Multiplying the first
condition by $\not n$ and the second by $\not\! m^{*}$ we obtain the
more conventional-looking conditions

\begin{equation}
\label{eq:nullconditions2}
  \begin{array}{rcl}
\not n \not l \epsilon^{I} & = & (1-\gamma^{uv})\epsilon^{I} = 0\, ,\\
& & \\
\not\! m^{*} \not\! m \epsilon^{I} & = & 
-(1+\gamma^{zz^{*}})\epsilon^{I} = 0\, .\\ 
  \end{array}
\end{equation}

\noindent
If $\epsilon^{I}$ satisfies the second condition, using
$\gamma_{5}=-\gamma^{uv}\gamma^{zz^{*}}$ 

\begin{equation}
\gamma^{zz^{*}}\epsilon^{I}= -\epsilon^{I}\, ,\,\, \Rightarrow\,\,\,   
\gamma^{uv}\gamma^{zz^{*}}\epsilon^{I}= \gamma^{uv}\epsilon^{I}\, ,\,\,
\Rightarrow -\gamma^{5}\epsilon^{I}= \gamma^{uv}\epsilon^{I}\, ,
\end{equation}

\noindent
which, due to the chirality of
$\epsilon^{I}$, leads to the first condition.

Let us now consider the KSE $\delta_{\epsilon}\psi_{I\, a}=0$. Taking
into account Eqs.~(\ref{eq:nullconditions}), our tetrad choice and
Eq.~(\ref{eq:doQ}), we find that the Killing spinors $\epsilon_{I}$
must be independent of $v,z,z^{*}$ and must satisfy 

\begin{equation}
\dot{\epsilon}_{I} 
+{\textstyle\frac{1}{2}}\epsilon_{IJ}\phi \gamma^{z^{*}}\epsilon^{J}=0\, ,
\end{equation}

\noindent
where $\phi=\mathcal{T}_{\Lambda}\phi^{\Lambda}=\phi(u)$.  
Observe that this equation can always be integrated, even though
the explicit form of the $\epsilon_{I}$ may be hard to find.

If $\phi (u)$ is a real function, however, the general solution is readily
found to be

\begin{equation}
\epsilon_{I}= e^{i\Phi} \epsilon_{I\, 0} 
+{\textstyle\frac{1}{\sqrt{2}}}\epsilon_{IJ}
\gamma^{z^{*}} e^{-i\Phi} \epsilon^{J}{}_{0}, ,
\end{equation}

\noindent
where

\begin{equation}
\gamma^{z^{*}} \epsilon_{I\, 0} = \gamma^{u}\epsilon_{I\, 0} =0\, ,
\hspace{1cm}
(\epsilon_{I\, 0})^{*}=\epsilon^{I}{}_{0} 
\hspace{1cm}
\dot{\Phi}=-i\phi/\sqrt{2}\, .
\end{equation}

Thus, all the configurations identified are supersymmetric and
preserve, at least $1/2$ of the available supersymmetries.
One can see, moreover, that the only configurations that preserve
more than $1/2$ are in fact maximally supersymmetric: 
Minkowski space and the maximally supersymmetric
wave of minimal $N=2$ $D=4$ supergravity found by Kowalski-Glikman
\cite{Kowalski-Glikman:1985im}, embedded such that only the graviphoton
is non-trivial.

\subsection{Equations of motion}
\label{sec:NullEOM}
Let us start with the Maxwell equations and Bianchi identities, given
in Eq.~(\ref{eq:MBnull}). There is only one non-trivial component
which is not automatically satisfied for supersymmetric configurations,
namely Eq.~(\ref{eq:MBnull2}), and we can rewrite it as

\begin{equation}
\label{eq:MBnull3}
e^{\mathcal{K}/2}\mathfrak{D}_{\underline{z}}(e^{-\mathcal{K}/2}\psi^{i})
\mathcal{U}_{i} +\psi^{i}\partial_{\underline{z}}Z^{j}\mathfrak{D}_{j}\mathcal{U}_{i}
-{\rm c.c.}=0\, ,
\end{equation}

\noindent
where one should keep in mind that the combination $e^{-\mathcal{K}/2}\psi^{i}$ is 
a weight $-1$ vector field.
Taking the symplectic product with $\mathcal{U}_{k}$ and using
Eqs.~(\ref{eq:SGProp1},\ref{eq:UU}) and (\ref{eq:SGDefC}), one finds

\begin{equation}
\label{eq:QueHorror}
\mathfrak{D}_{\underline{z}^{*}}(e^{-\mathcal{K}/2}\psi^{*\, i^{*}})
-ie^{-\mathcal{K}/2}\psi^{j}\partial_{\underline{z}}Z^{k}
\mathcal{C}_{jk}{}^{i^{*}}=0\, .
\end{equation}

A somewhat lighter equation can be derived by defining 
\begin{equation}
  \label{eq:QueHorror2}
  \psi^{i} \ =\ e^{\mathcal{K}}\mathcal{G}^{ij^{*}}\ P_{j^{*}} \;\;\rightarrow\;\;
  \partial_{\underline{z}^{*}}P^{*}_{i} \ =\ i\mathcal{C}_{ij}{}^{k^{*}}\ \partial_{z}Z^{j}\ P_{k^{*}} \; ,
\end{equation}
where $P_{i^{*}}$ is of K\"ahler weight $(0,2)$.
This equation determines $\psi^{i}$, but it is extremely difficult to
find a general solution, although we will give some solutions in
Appendix~\ref{appsec:SpecShit}.

The only non-automatically satisfied component of the Einstein equations
is the $uu$ one

\begin{equation}
\mathcal{E}_{uu}= R_{uu} +2 \mathcal{G}_{ij^{*}}A^{i}A^{*\, j^{*}}
-i\Im{\rm m}\mathcal{N}_{\Lambda\Sigma} \phi^{\Lambda}\phi^{\Sigma} =0\, .
\end{equation}

Using Eq.~(\ref{eq:inverse}), and the value of $R_{uu}$ this equation
takes the form

\begin{equation}
\label{eq:EEuu}
  \begin{array}{rcl}
-2 e^{-2U}\partial_{\underline{z}}\partial_{\underline{z}^{*}}H
+{\textstyle\frac{1}{2}} e^{-4U}(\partial_{\underline{z}^{*}}\omega_{\underline{z}} 
-\partial_{\underline{z}}\omega_{\underline{z}^{*}})^{2}  
+e^{-2U}(\partial_{\underline{z}^{*}}\dot{\omega}_{\underline{z}} 
+\partial_{\underline{z}}\dot{\omega}_{\underline{z}^{*}})& & \\
& & \\
+2(\ddot{U}+\dot{U}\dot{U}) +2\mathcal{G}_{ij^{*}}(A^{i}A^{*\, j^{*}} 
+8\psi^{i}\psi^{*\, j^{*}}) +4|\phi|^{2} & = & 0\, .\\
\end{array}
\end{equation}

A supersymmetric solution in this class is, then, fully determined by
the real function $H(z,z^{*},u)$ and the complex functions
$\omega_{\underline{z}}(z,z^{*},u),\phi(u),\psi^{i}(z,z^{*},u),Z^{i}(z,u)$
satisfying Eqs.~(\ref{eq:DetOmegaQ},\ref{eq:MBnull3}) and (\ref{eq:EEuu}). There are two
simple and interesting families of solutions

\begin{enumerate}
\item $Z^{i}=Z^{i}(z)$. ($A^{i}=0$). This implies that
  $\mathcal{Q}_{\underline{u}}=0$ and we can safely take
  $\hat{\omega}=0$. The Einstein equation takes the form

\begin{equation}
 e^{\mathcal{K}}\partial_{\underline{z}}\partial_{\underline{z}^{*}}H
=8\mathcal{G}_{ij^{*}}\psi^{i}\psi^{*\, j^{*}} +2|\phi|^{2}\, ,
\end{equation}

\noindent
and can be integrated once the solutions to the Maxwell and Bianchi
equations, $\psi^{i}$, are given ($\phi(u)$ is an arbitrary
complex function).

If we set to zero the vector field strengths $\phi=\psi^{i}=0$, the
Einstein equation reduces to the statement that $H$ is a real harmonic function on $\mathbb{C}$.

The solutions in this subclass are combinations of $pp$-waves associated
to the harmonic function $H$ and cosmic strings of the kind considered
in Ref.~\cite{Greene:1989ya}, {\em i.e.\/} determined by $n$
holomorphic functions $Z^{i}=Z^{i}(z)$. Now the metric is determined
by supersymmetry to be

\begin{equation}
\left\{
  \begin{array}{rcl}
ds^{2} & = & 2du(dv+Hdu) -2e^{-2\mathcal{K}(Z,Z^{*})}dzdz^{*}\, ,\\
& & \\
Z^{i} & = & Z^{i}(z)\, ,\\
& & \\
\partial_{\underline{z}}\partial_{\underline{z}^{*}}H & = & 0\, .\\
  \end{array}
\right.
\end{equation}

In order to study the behaviour of these solutions under the symmetries of the
theory, it is convenient to express them in an arbitrary system of
holomorphic coordinates

\begin{equation}
\left\{
  \begin{array}{rcl}
ds^{2} & = & 2du(dv+Hdu) -2e^{-[\mathcal{K}(Z,Z^{*})-h-h^{*}]} dzdz^{*}\, ,\\
& & \\
Z^{i} & = & Z^{i}(z)\, ,\\
& & \\
h & = & h(z)\, ,\\
& & \\
\partial_{\underline{z}}\partial_{\underline{z}^{*}}H & = & 0\, .\\
  \end{array}
\right.
\end{equation}

The Killing spinors of these solutions are 

\begin{equation}
\epsilon_{I}= e^{-\frac{1}{4}(h-h^{*})}\epsilon_{I\, 0}\, ,
\hspace{1cm}
\gamma^{z^{*}}\epsilon_{I\, 0}=0\, .
\end{equation}

The isometries of the K\"ahler metric (which are the duality
symmetries of our theory) leave invariant the K\"ahler potential up to
K\"ahler transformations Eq.~(\ref{eq:Kpotentialtransformation}). Of
course, these duality transformations leave invariant the spacetime
metric, but the relation between the $g_{zz^{*}}$ component and the
K\"ahler potential will change unless the holomorphic function $h$
transforms according to

\begin{equation}
h^{\prime} = h +f\, ,  
\end{equation}

\noindent
which makes the Killing spinors transform precisely as objects of
K\"ahler weight $1/2$, as they should. Actually, for the metric to be
form-invariant, it is enough that $\Re{\rm e} (h^{\prime}) = \Re{\rm
  e} (h^{\prime})+ \Re{\rm e}(f)$ while the spinors will behave as
objects of K\"ahler weight $1/2$ if $\Im{\rm m} (h^{\prime}) = \Im{\rm
  m} (h^{\prime})+ \Im{\rm m}(f)$. These two conditions are
independent. Only the first was required in the construction of
Ref.~\cite{Greene:1989ya}, but supersymmetry requires the
second.\footnote{We thank Jelle Hartong for useful conversations on
  this point.}

The holomorphic functions $Z^{i}(z)$ will in general be multi-valued
and will have non-trivial monodromies. Only those which are isometries
of the K\"ahler metric can be allowed. The metric will be invariant
and the Killing spinors will also have the correct monodromy if $h$
transforms as above.

\item $Z^{i}=Z^{i}(u)= 0$. This implies that $\mathcal{K}$ and,
  therefore, $U$ are functions of $u$ only, whence the latter can be
  eliminated from the metric by a change of coordinates.  Since the
  pullback of K\"ahler 1-form depends on $u$ only, we can solve
  Eq.~(\ref{eq:doQ}) for $\hat{\omega}$:

  \begin{equation}
  \hat{\omega} =i e^{-\mathcal{K}}\mathcal{Q}_{\underline{u}}
(zdz^{*}-z^{*}dz)\, ,  
  \end{equation}
  
\noindent
which can, however, be eliminated by further change of coordinates.
The remaining Einstein equation takes the form

\begin{equation}
2 \partial_{\underline{z}}\partial_{\underline{z}^{*}}H =
 2\mathcal{G}_{ij^{*}}(A^{i}A^{*\, j^{*}}+8\psi^{i}\psi^{*\, j^{*}}) +4|\phi|^{2} \, .
\end{equation}

\noindent
Eq.~(\ref{eq:QueHorror}) can in this case be solved, leading to
the statement that $\psi^{i}$ only depends on $u$ and $z^{*}$.
Introducing then the functions $\Upsilon^{i}$, defined through the 
relation $\partial_{z^{*}}\Upsilon^{i} =\psi^{i}$, the above equation
can be integrated with great ease, giving 

\begin{equation}
H =(
\mathcal{G}_{ij^{*}}\dot{Z}^{i}\dot{Z}^{*\, j^{*}} +2|\phi|^{2})|z|^{2} 
+8\mathcal{G}_{ij^{*}}\Upsilon^{i}\Upsilon^{*}{}^{j^{*}}+f(z,u) +f^{*}(z^{*},u)\, .
\end{equation}

The supersymmetric solutions of this class take, therefore, the form

\begin{equation}
\left\{
  \begin{array}{rcl}
ds^{2} & = & 2du(dv+Hdu) -2dzdz^{*}\, ,\\
& & \\
F^{\Lambda\, +} & = &  
  \left[ \frac{i}{2}\mathcal{L}^{*\, \Lambda}\phi(u)
     + f^{\Lambda}_{i}\psi^{i}(u,z^{*})
  \right] du\wedge dz^{*}\, ,\\
& & \\
Z^{i} & = & Z^{i}(u)\, ,\\
  \end{array}
\right.
\end{equation}

\noindent
where $Z^{i},\phi$ are arbitrary functions of $u$ and $H$ is given
above.

\end{enumerate}


\section*{Acknowledgments}

T.O.~would like to thank interesting conversations with I.~Bena and
R.~Kallosh and the PH-TH Division of CERN for their hospitality and
financial support during the last stages of this work and
M.M.~Fern\'andez for her lasting support.  Likewise, PM thanks the
Instituto de Física Teórica for its hospitality. This work has been
supported in part by the Spanish grant BFM2003-01090.

\appendix

\section{Conventions}
\label{sec-conventions}

In this paper we use basically the notation of Ref.~\cite{Andrianopoli:1996cm}
and the conventions of Ref.~\cite{Bellorin:2005zc}, to which we have adapted
the formulae of Ref.~\cite{Andrianopoli:1996cm}. The main differences between
the conventions of those two references are the signs of spin connection, the
completely antisymmetric tensor $\epsilon^{abcd}$ and $\gamma_{5}$. Thus,
chiralities are reversed and self-dual tensors are replaced by anti-self-dual
tensors and vice-versa. The curvatures are identical. Finally, the
normalization of the 2-form components differs by a factor of 2: for us

\begin{equation}
F=dA={\textstyle\frac{1}{2}} F_{\mu\nu}dx^{\mu}\wedge dx^{\nu} \Rightarrow
F_{\mu\nu} = 2\partial_{[\mu}A_{\nu]}\, ,
\end{equation}

\noindent
which amounts to a difference of a factor of 2 in the vectors supersymmetry
transformations Eq.~(\ref{eq:susytransA}). Further, all fermions and
supersymmetry parameters from Ref.~\cite{Andrianopoli:1996cm} have been
rescaled by a factor of $\frac{1}{2}$, which introduces additional factors of
$\frac{1}{4}$ in all the bosonic fields supersymmetry transformations
Eqs.~(\ref{eq:susytranse}-\ref{eq:susytransZ}).

The meaning of the different indices used in this paper is explained in
Table~\ref{tab-indices}.  We use the shorthand $\bar{n}\equiv n+1$.

\begin{table}[ht]
\centering
\begin{tabular}{|l|l|}
\hline
 Type & Associated structure \\
\hline\hline
 $\mu$, $\nu$, $\ldots$  &  Curved space \\
\hline
 $a$, $b$, $\ldots$  &  Tangent space \\
\hline
 $m,n,\ldots$ & Cartesian $\mathbb{R}^{3}$-indices\\
\hline
 $i,j,\ldots$; $i^{*},j^{*},\ldots$ & Complex scalar fields and 
their conjugates. There are $n$ of them.\\
\hline
 $\Lambda ,\Sigma ,\ldots$ & $\mathfrak{sp}(\bar{n})$ indices ($\bar{n}=n+1$)\\
\hline
 $I,J ,\ldots$ & $N=2$ spinor indices \\
\hline
\end{tabular}
\caption{Meaning of the indices used in this paper.}

\label{tab-indices}
\end{table}


\section{K\"ahler geometry}
\label{sec-kahlergeometry}

A K\"ahler manifold $\mathcal{M}$ is a complex manifold on which there
exist complex coordinates $z^{i}$ and $z^{*\, i^{*}} = (z^{i})^{*}$
and a function $\mathcal{K}$, called the {\em K\"ahler potential},
such that the line element is

\begin{equation}
ds^{2} = 2 \mathcal{G}_{ii^{*}}\ dz^{i}dz^{*\, i^{*}}\, ,
\end{equation}

\noindent
with

\begin{equation}
\label{eq:Kmetric}
\mathcal{G}_{ii^{*}} = \partial_{i}\partial_{i^{*}}\mathcal{K}\, .
\end{equation}

The \textit{K\"ahler (connection) 1-form}  $\mathcal{Q}$ is defined by

\begin{equation}
\label{eq:K1form}
\mathcal{Q} \equiv (2i)^{-1}(dz^{i}\partial_{i}\mathcal{K} -
dz^{*\, i^{*}}\partial_{i^{*}}\mathcal{K})\, ,
\end{equation}

\noindent
and the \textit{K\"ahler 2-form} $\mathcal{J}$ is its exterior
derivative

\begin{equation}
\label{eq:K2form}
\mathcal{J} \equiv d\mathcal{Q} = i\mathcal{G}_{ii^{*}} 
dz^{i}\wedge dz^{*\, i^{*}}\, .
\end{equation}

The Levi-Civit\`a connection on a K\"ahler manifold is given by

\begin{equation}
\label{eq:KCCChrisSymb}
\Gamma_{jk}{}^{i} = 
\mathcal{G}^{ii^{*}}\partial_{j}\mathcal{G}_{i^{*}k}\, ,
\hspace{1cm}
\Gamma_{j^{*}k^{*}}{}^{i^{*}}  = 
\mathcal{G}^{i^{*}i}\partial_{j^{*}}\mathcal{G}_{k^{*}i} \, .\end{equation}

\noindent
The Riemann curvature tensor has as only non-vanishing components
$R_{ij^{*}kl^{*}}$, but we will not need their explicit expression.  The Ricci
tensor is given by

\begin{equation}
\label{eq:KCCRicci}
R_{ii^{*}}  = \partial_{i}\partial_{i^{*}}
\left(\textstyle{1\over 2}\log\det\mathcal{G}\right) \, .
\end{equation}

The K\"ahler potential is not unique: it is defined only up to
\textit{K\"ahler transformations} of the form

\begin{equation}
\label{eq:Kpotentialtransformation}
\mathcal{K}^{\prime}(z,z^{*})=\mathcal{K}(z,z^{*})+f+f^{*}\, , 
\end{equation}

\noindent
where $f$ is any holomorphic function of the complex coordinates
$z^{i}$. Under these transformations, the K\"ahler metric and K\"ahler
2-form are invariant, while the components of the K\"ahler connection
1-form transform according to

\begin{equation}
\label{eq:K1formtransformation}
\mathcal{Q}^{\prime}_{i} =\mathcal{Q}_{i} 
-{\textstyle\frac{i}{2}}\partial_{i}f\, .
\end{equation}

By definition, objects with K\"ahler weight $(q,\bar{q})$ transform
under the above K\"ahler transformations with a factor
$e^{-(qf+\bar{q}f^{*})/2}$ and the K\"ahler-covariant derivative
$\mathfrak{D}$ acting on them is given by

\begin{equation}
\label{eq:Kcovariantderivative}
\mathfrak{D}_{i} \equiv \nabla_{i} +iq \mathcal{Q}_{i}\, ,
\hspace{1cm}
\mathfrak{D}_{i^{*}} \equiv \nabla_{i^{*}} -i\bar{q} \mathcal{Q}_{i^{*}}\, ,
\end{equation}

\noindent
where $\nabla$ is the standard covariant derivative
  associated to the Levi-Civit\`a connection on $\mathcal{M}$.
  
  When $(q,\bar{q})=(1,-1)$, this defines a complex line bundle
  $L^{1}\rightarrow \mathcal{M}$ over the K\"ahler manifold
  $\mathcal{M}$ whose first, and only, Chern class equals the K\"ahler
  2-form $\mathcal{J}$. A complex line bundle with this property is
  known as a \textit{K\"ahler-Hodge (KH) manifold} and provides the
  formal starting point for the definition of a special K\"ahler
  manifold\footnote{Some basic references for this material are
    \cite{Ceresole:1995ca,Ceresole:1995jg,Craps:1997gp} and the review
    \cite{kn:toinereview}. The definition of special K\"ahler manifold
    was made in Ref.~\cite{Strominger:1990pd}, formalizing the
    original results of Ref.~\cite{deWit:1984px}.} that is explained
  in the next Appendix.

We will often use the spacetime pullback of the K\"ahler-covariant
derivative on tensor fields with K\"ahler weight $(q,-q)$ (weight $q$, for
short) for which it takes the simple form

\begin{equation}
\label{eq:Kcovariantderivative2}
\mathfrak{D}_{\mu}= \nabla_{\mu} +iq\mathcal{Q}_{\mu}\, ,
\end{equation}

\noindent
where $\nabla_{\mu}$ is the standard spacetime covariant derivative plus
possibly the pullback of the Levi-Civit\`a connection on $\mathcal{M}$; $\mathcal{Q}_{\mu}$ is
the pullback of the K\"ahler 1-form, {\em i.e.}

\begin{equation}
\label{eq:KahlerConPB}
\mathcal{Q}_{\mu} =  (2i)^{-1}(\partial_{\mu}z^{i}\partial_{i}\mathcal{K} -
\partial_{\mu}z^{*\, i^{*}}\partial_{i^{*}}\mathcal{K})\, .
\end{equation}


\section{Special K\"ahler geometry}
\label{sec-specialgeometry}

Let us now consider a flat $2\bar{n}$-dimensional vector bundle
$E\rightarrow\mathcal{M}$ with structure group
$Sp(\bar{n};\mathbb{R})$, and take a section $\mathcal{V}$ of the
product bundle $E\otimes L^{1}\rightarrow\mathcal{M}$ and its complex
conjugate $\overline{\mathcal{V}}$, which formally is a section of the
bundle $E\otimes L^{-1}\rightarrow \mathcal{M}$. Then, a
\textit{special K\"ahler manifold}, is a bundle $E\otimes
L^{1}\rightarrow\mathcal{M}$, for which there exists a section
$\mathcal{V}$ such that

\begin{equation}
\label{eq:SGDefFund}
\mathcal{V} = 
\left(
\begin{array}{c}
\mathcal{L}^{\Lambda}\\
\mathcal{M}_{\Sigma}\\
\end{array}
\right) \;\; \rightarrow \;\;
\left\{
\begin{array}{lcl}
\langle \mathcal{V}\mid\mathcal{V}^{*}\rangle 
& \equiv & 
\mathcal{L}^{*\, \Lambda}\mathcal{M}_{\Lambda} 
-\mathcal{L}^{\Lambda}\mathcal{M}^{*}_{\Lambda}
= -i\, , \\
& & \\
\mathfrak{D}_{i^{*}}\mathcal{V} & = & (\partial_{i^{*}}+
{\textstyle\frac{1}{2}}\partial_{i^{*}}\mathcal{K})\mathcal{V} =0 \, ,\\
& & \\
\langle\mathfrak{D}_{i}\mathcal{V}\mid\mathcal{V}\rangle & = & 0 \, .
\end{array}
\right. 
\end{equation}

If we then define

\begin{equation}
\label{eq:SGDefU}
\mathcal{U}_{i}  \equiv \mathfrak{D}_{i}\mathcal{V}  
= 
\left(
\begin{array}{c}
f^{\Lambda}{}_{i}\\
h_{\scriptscriptstyle{\Sigma}\, i}
\end{array}
\right)\, ,\,\,\,\, 
\mathcal{U}^{*}{}_{i^{*}} = (\mathcal{U}_{i})^{*} \, ,
\end{equation}

\noindent
then it follows from the basic definitions that

\begin{equation}
\label{eq:SGProp1}
\begin{array}{rclrcl}
\mathfrak{D}_{i^{*}}\ \mathcal{U}_{i} 
& = & 
\mathcal{G}_{ii^{*}}\ \mathcal{V}\,\hspace{2cm} &
\langle\mathcal{U}_{i}\mid\mathcal{U}^{*}{}_{i^{*}}\rangle 
& = & 
i\mathcal{G}_{ii^{*}} \, , \\
& & & & & \\
\langle\mathcal{U}_{i}\mid\mathcal{V}^{*}\rangle &  = & 0\, , &
\langle\mathcal{U}_{i}\mid\mathcal{V}\rangle & = & 0 \, .\\
\end{array}
\end{equation}

Taking the covariant derivative of the last identity
$\langle\mathcal{U}_{i}\mid\mathcal{V}\rangle =0$ we find immediately that
$\langle\mathfrak{D}_{i}\mathcal{U}_{j}\mid\mathcal{V}\rangle = -\langle\ 
\mathcal{U}_{j}\mid \mathcal{U}_{i}\rangle$. It can be shown that the
r.h.s.~of this equation is antisymmetric while the l.h.s.~is symmetric, so that

\begin{equation}
\label{eq:UU}
\langle\mathfrak{D}_{i}\mathcal{U}_{j}\mid\mathcal{V}\rangle = \langle
\mathcal{U}_{j}\mid \mathcal{U}_{i}\rangle = 0\, .
\end{equation}

The importance of this last equation is that if we group together
$\mathcal{E}_{\Lambda} = (\mathcal{V},\mathcal{U}_{i})$, we can see that
$\langle \mathcal{E}_{\Sigma}\mid\mathcal{E}^{*}{}_{\Lambda}\rangle$ is a
non-degenerate matrix. This then allows us to construct an identity operator
for the symplectic indices, such that for a given section of
$\mathcal{A}\ni\Gamma\left( E,\mathcal{M}\right)$ we have

\begin{equation}
\label{eq:SGSymplProj}
\mathcal{A} = i\langle\mathcal{A}\mid\mathcal{V}^{*}\rangle \mathcal{V} 
-i\langle\mathcal{A}\mid\mathcal{V}\rangle\ \mathcal{V}^{*}
+i\langle\mathcal{A}\mid\mathcal{U}_{i}\rangle\mathcal{G}^{ii^{*}}\ 
\mathcal{U}^{*}{}_{i^{*}}
-i\langle\mathcal{A}\mid\mathcal{U}^{*}{}_{i^{*}}\rangle
\mathcal{G}^{ii^{*}}\mathcal{U}_{i} \, .
\end{equation}

As we have seen $\mathfrak{D}_{i}\mathcal{U}_{j}$ is symmetric in $i$ and $j$,
but what more can be said about it: as one can easily see, the inner product
with $\mathcal{V}^{*}$ and $\mathcal{U}^{*}{}_{i^{*}}$ vanishes due to the
basic properties. Let us then define the K\"ahler-weight 2 object

\begin{equation}
\label{eq:SGDefC}
\mathcal{C}_{ijk} \equiv 
\langle \mathfrak{D}_{i}\ \mathcal{U}_{j}\mid \mathcal{U}_{k}\rangle 
\;\; \rightarrow\;\; 
\mathfrak{D}_{i}\ \mathcal{U}_{j}  = 
i\mathcal{C}_{ijk}\mathcal{G}^{kl^{*}}\mathcal{U}^{*}{}_{l^{*}} \, ,
\end{equation}

\noindent
where the last equation is a consequence of
Eq.~(\ref{eq:SGSymplProj}). Since the $\mathcal{U}$'s are orthogonal,
however, one can see that $\mathcal{C}$ is completely symmetric in its
3 indices. Furthermore one can show that

\begin{equation}
\label{eq:SGCProp}
\mathfrak{D}_{i^{*}}\ \mathcal{C}_{jkl}  = 0\, ,\hspace{1cm}
\mathfrak{D}_{[i}\ \mathcal{C}_{j]kl} = 0\, .
\end{equation}

Observe that these equations imply the existence of a function $\mathcal{S}$,
such that

\begin{equation}
\mathcal{C}_{ijk} =
\mathfrak{D}_{i}\mathfrak{D}_{j}\mathfrak{D}_{k}\ \mathcal{S}\, .
\end{equation}

The function $\mathcal{S}$ is given by \cite{Castellani:1990zd} 

\begin{equation}
\mathcal{S}\sim
\mathcal{L}^{\Lambda}
\Im{\rm m}\mathcal{N}_{\Lambda\Sigma}\mathcal{L}^{\Sigma}\, ,
\end{equation}

\noindent
where $\mathcal{N}$ is the period or monodromy matrix. This matrix is defined
by the relations

\begin{equation}
\label{eq:SGDefN}
\mathcal{M}_{\Lambda}  = \mathcal{N}_{\Lambda\Sigma} \mathcal{L}^{\Sigma}\, ,
\hspace{1cm}
h_{\Lambda\, i}  = \mathcal{N}^{*}{}_{\Lambda\Sigma} f^{\Sigma}{}_{i} \, .
\end{equation}

\noindent
The relation $\langle\mathcal{U}_{i}\mid\overline{\mathcal{V}}\rangle =0$
then implies that $\mathcal{N}$ is symmetric, which then also trivializes
$\langle\mathcal{U}_{i}\mid\mathcal{U}_{j}\rangle =0$.

From the other basic properties in (\ref{eq:SGProp1}) we find

\begin{eqnarray}
\mathcal{L}^{\Lambda} \Im{\rm m}\mathcal{N}_{\Lambda\Sigma}
\mathcal{L}^{*\, \Sigma} & = & -{\textstyle\frac{1}{2}}\, ,
\label{eq:esta}\\
& & \nonumber \\
\mathcal{L}^{\Lambda} \Im{\rm m}\mathcal{N}_{\Lambda\Sigma} f^{\Sigma}{}_{i} 
& = & 
\mathcal{L}^{\Lambda} \Im{\rm m}\mathcal{N}_{\Lambda\Sigma} 
f^{*\, \Sigma}{}_{i^{*}} 
=0\, ,
\label{eq:esa}\\
& & \nonumber \\
f^{\Lambda}{}_{i}\ \Im{\rm m}\mathcal{N}_{\Lambda\Sigma}
f^{*\, \Sigma}{}_{i^{*}} & = &   -\textstyle{1\over 2}\mathcal{G}_{ii^{*}} \, .
\label{eq:esaotra}
\end{eqnarray}

Further identities that can be derived are

\begin{eqnarray}
\label{eq:SGColl1}
(\partial_{i}\mathcal{N}_{\Lambda\Sigma}) \mathcal{L}^{\Sigma} 
& = & 
-2i\Im{\rm m}(\mathcal{N})_{\Lambda\Sigma}\ f^{\Sigma}{}_{i} \, , \\
& & \nonumber \\
\label{eq:SGColl2}
\partial_{i}\mathcal{N}^{*}{}_{\Lambda\Sigma}\ f^{\Sigma}{}_{j} 
& = & 
-2\mathcal{C}_{ijk}\mathcal{G}^{kk^{*}}
\Im{\rm m}\mathcal{N}_{\Lambda\Sigma}
f^{*\, \Sigma}{}_{k^{*}} \, ,\\
& & \nonumber \\
\label{eq:SGColl3}
\mathcal{C}_{ijk} 
& = & 
f^{\Lambda}{}_{i}f^{\Sigma}{}_{j} 
\partial_{k}\mathcal{N}^{*}_{\Lambda\Sigma} \, ,\\
& & \nonumber \\
\label{eq:SGColl4}
\mathcal{L}^{\Sigma}\partial_{i^{*}}\mathcal{N}_{\Lambda\Sigma} 
& = & 
0 \, , \\
& & \nonumber \\
\label{eq:SGColl5}
\partial_{i^{*}}\mathcal{N}^{*}{}_{\Lambda\Sigma}\ f^{\Sigma}{}_{i} 
& = & 
2i\mathcal{G}_{ii^{*}}\Im{\rm m}\mathcal{N}_{\Lambda\Sigma}
 \mathcal{L}^{\Sigma} \, .
\end{eqnarray}

An important identity one can derive, and that will be used various times
in the main text, is given by

\begin{equation}
\label{eq:SGImpId}
U^{\Lambda\Sigma} \equiv  f^{\Lambda}{}_{i}\mathcal{G}^{ii^{*}}
f^{*\, \Sigma}{}_{i^{*}}
=
-\textstyle{1\over 2}\Im{\rm m}(\mathcal{N})^{-1|\Lambda\Sigma}
-\mathcal{L}^{*\, \Lambda}\mathcal{L}^{\Sigma} \; ,
\end{equation}

\noindent
whence $(U^{\Lambda\Sigma})^{*}=U^{\Sigma\Lambda}$.  




We can define the graviphoton and matter vector projectors

\begin{eqnarray}
\label{eq:projectorg}
\mathcal{T}_{\Lambda} & \equiv & 2i \mathcal{L}_{\Lambda}
=2i\mathcal{L}^{\Sigma}\Im{\rm m}\, 
\mathcal{N}_{\Sigma\Lambda}\, ,\\
& & \nonumber \\
\label{eq:projectorm}
\mathcal{T}^{i}{}_{\Lambda} & \equiv & -f^{*}{}_{\Lambda}{}^{i}=
-\mathcal{G}^{ij^{*}}
f^{*\, \Sigma}{}_{j^{*}}\Im{\rm m}\, 
\mathcal{N}_{\Sigma\Lambda}\, .
\end{eqnarray}

Using these definitions and the above properties one can show the following
identities for the derivatives of the period matrix:

\begin{equation}
  \label{eq:dN}
  \begin{array}{rcl}
\partial_{i}\mathcal{N}_{\Lambda\Sigma} & = & 
4\mathcal{T}_{i(\Lambda}\mathcal{T}_{\Sigma)}\, ,\\
& & \\
\partial_{i^{*}}\mathcal{N}_{\Lambda\Sigma} & = & 
4\mathcal{C}^{*}{}_{i^{*} j^{*} k^{*}}
\mathcal{T}^{i^{*}}{}_{(\Lambda}\mathcal{T}^{j^{*}}{}_{\Sigma)}\, .\\
  \end{array}
\end{equation}


\subsection{Prepotential: Existence and more formulae}
\label{subsec:Prepot}

Let us start by introducing the explicitly holomorphic section $\Omega =
e^{-\mathcal{K}/2}\mathcal{V}$, which allows us to rewrite the system
Eqs.~(\ref{eq:SGDefFund}) as

\begin{equation}
\label{eq:SGDefFund2}
\Omega =
\left(
\begin{array}{c}
\mathcal{X}^{\Lambda}\\
\mathcal{F}_{\Sigma}\\
\end{array}
\right) 
\;\; \rightarrow \;\;
\left\{
\begin{array}{lcl}
\langle \Omega \mid\Omega^{*}\rangle 
& \equiv & 
\mathcal{X}^{*\, \Lambda}\mathcal{F}_{\Lambda}  
-\mathcal{X}^{\Lambda}\mathcal{F}^{*}_{\Lambda}
= -i\ e^{-\mathcal{K}}\, , \\
& & \\
\partial_{i^{*}}\Omega & = & 0 \, ,\\
& & \\
\langle\partial_{i}\Omega\mid\Omega\rangle & = & 0 \; .
     \end{array}
  \right. 
\end{equation}

Observe that the first of Eqs.~(\ref{eq:SGDefFund2}) together with the
definition of the period matrix $\mathcal{N}$ imply the following
expression for the K\"ahler potential: 

\begin{equation}
\label{eq:kpotential}
e^{-\mathcal{K}}= -2\Im{\rm m}\mathcal{N}_{\Lambda\Sigma}
\mathcal{X}^{\Lambda}  \mathcal{X}^{*\, \Sigma}\, .
\end{equation}

If we now assume that $\mathcal{F}_{\Lambda}$ depends on $Z^{i}$ through the
$\mathcal{X}$'s, then from the last equation we can derive that

\begin{equation}
\partial_{i}\mathcal{X}^{\Lambda}
\left[ 
2\mathcal{F}_{\Lambda}  
-\partial_{\Lambda}\left( \mathcal{X}^{\Sigma}\mathcal{F}_{\Sigma}\right)
\right] 
= 0 \; .
\end{equation}

If $\partial_{i}\mathcal{X}^{\Lambda}$ is invertible as an $n\times \bar{n}$
matrix, then we must conclude that
 
\begin{equation}
\label{eq:Prepot}
\mathcal{F}_{\Lambda}  = \partial_{\Lambda}\mathcal{F}(\mathcal{X}) \, ,
\end{equation}

\noindent
where $\mathcal{F}$ is a homogeneous function of degree 2, called the
\textit{prepotential}.

Making use of the prepotential and the definitions (\ref{eq:SGDefN}), we
can calculate

\begin{equation}
\label{eq:PrepotN}
\mathcal{N}_{\Lambda\Sigma} 
=\mathcal{F}^{*}_{\Lambda\Sigma} 
+2i\frac{
\Im{\rm m}\mathcal{F}_{\Lambda\Lambda^{\prime}}
\mathcal{X}^{\Lambda^{\prime}}
\Im{\rm m}\mathcal{F}_{\Sigma\Sigma^{\prime}}\mathcal{X}^{\Sigma^{\prime}}
}
{                                          
\mathcal{X}^{\Omega}\Im{\rm m}\mathcal{F}_{\Omega\Omega^{\prime}}
\mathcal{X}^{\Omega^{\prime}}
}\, .
\end{equation}

Having the explicit form of $\mathcal{N}$, we can also derive an explicit
representation for $\mathcal{C}$ by applying Eq. (\ref{eq:SGColl4}). One finds

\begin{equation}
\label{eq:PrepC}
\mathcal{C}_{ijk} = 
e^{\mathcal{K}}\partial_{i}\mathcal{X}^{\Lambda}
 \partial_{j}\mathcal{X}^{\Sigma} \partial_{k}\mathcal{X}^{\Omega}
\mathcal{F}_{\Lambda\Sigma\Omega} \; ,
\end{equation}

\noindent
so that the prepotential really determines all structures in special geometry.

A last remark has to be made about the existence of a prepotential: clearly,
given a holomorphic section $\Omega$ a prepotential need not exist. It was
shown in Ref.~\cite{Craps:1997gp}, however, that one can always apply an
$Sp(\bar{n},\mathbb{R})$ transformation such that a prepotential exists.
Clearly the $N=2$ SUGRA action is not invariant under the full
$Sp(\bar{n},\mathbb{R})$, but the equations of motion and the supersymmetry
equations are. This means that for the purpose of this article we can always,
even if this is not done, impose the existence of a prepotential.


\section{Some explicit cases}
\label{appsec:SpecShit}


\subsection{Quadratic prepotential}

This is a simple, but important, case in which there is a prepotential
and it takes the form

\begin{equation}
\mathcal{F}= \textstyle{1\over 2}\mathcal{F}_{\Lambda\Sigma}
\mathcal{X}^{\Lambda}\mathcal{X}^{\Sigma}\, ,
\end{equation}

\noindent
where $\mathcal{F}_{\Lambda\Sigma}$ is a complex, symmetric, constant
matrix that coincides with the matrix of second derivatives of
$\mathcal{F}$. Its imaginary part must be negative definite.  The
period matrix is given by Eq.~(\ref{eq:PrepotN}).
Observe that

\begin{equation}
\label{eq:relation}
\mathcal{F}_{\Lambda} = \mathcal{F}_{\Lambda\Sigma}\mathcal{X}^{\Sigma}\, .
\end{equation}

The K\"ahler potential is 

\begin{equation}
\label{eq:Kahlerpotentialquadraticprepotential}
e^{-\mathcal{K}}= -2\Im{\rm m}\mathcal{N}_{\Lambda\Sigma}
\mathcal{X}^{*\, \Lambda}  
\mathcal{X}^{\Sigma}\, .  
\end{equation}

To construct the general solution of the timelike case, we need to
relate the real section $\mathcal{R}$ and $\mathcal{I}$ defined in
Eqs.~(\ref{eq:SGDefN}). This can be done by using 
the property Eq.~(\ref{eq:relation}), rescaling it by $e^{\mathcal{K}/2}/X$:

\begin{equation}
\mathcal{M}_{\Lambda}/X = \mathcal{F}_{\Lambda\Sigma}\mathcal{L}^{\Sigma}/X\, , 
\end{equation}

\noindent
and then, taking the imaginary part of this equation and using the
invertibility of $\Im{\rm m}\mathcal{F}_{\Lambda\Sigma}$, we find the
solution

\begin{equation}
\Re{\rm e} (\mathcal{L}^{\Lambda}/X) = \Im{\rm m}(\mathcal{F})^{-1|\Lambda\Sigma}
[\Im{\rm m}(\mathcal{M}_{\Sigma}/X) 
-\Re{\rm e}(\mathcal{F}_{\Sigma\Omega} \Im{\rm m}(\mathcal{L}^{\Omega}/X)]\, ,
\end{equation}

\noindent
which implies

\begin{equation}
\mathcal{L}^{\Lambda}/X = \Im{\rm m}(\mathcal{F})^{-1|\Lambda\Sigma}
[\Im{\rm m}(\mathcal{M}_{\Sigma}/X) 
-\mathcal{F}^{*}{}_{\Sigma\Omega})\Im{\rm m}(\mathcal{L}^{\Omega}/X)]\, .
\end{equation}

\noindent
In other words, this implies that we can take the components of the
section $\mathcal{L}^{\Lambda}/X$ to be arbitrary complex harmonic
functions $\mathcal{H}^{\Lambda}$. 

Now, $|M|^{2}$ which appears in the metric Eq.~(\ref{eq:metric}) is
given, according to Eq.~(\ref{eq:Msquared}), by the K\"ahler potential
Eq.~(\ref{eq:Kahlerpotentialquadraticprepotential}) where the scalars
$\mathcal{X}^{\Lambda}$ are substituted by the complex harmonic
functions $\mathcal{H}^{\Lambda}$, i.e.

\begin{equation}
|M|^{2}=2|X|^{2}= -[\Im{\rm m}\mathcal{F}_{\Lambda\Sigma}
\mathcal{H}^{*\, \Lambda}  \mathcal{H}^{\Sigma}]^{-1}\, .
\end{equation}

The other term that appears in the metric $\omega$, is given in terms
of the real section $\mathcal{I}$ by Eq.~(\ref{eq:oidi}). Substituting
the imaginary parts of the harmonic functions $ \mathcal{H}^{\Lambda}$
in that formula, we get 

\begin{equation}
(d\omega)_{mn} =
-i\epsilon_{mnp}\Im{\rm m}\mathcal{F}_{\Lambda\Sigma}
[\partial_{p}\mathcal{H}^{\Lambda} \mathcal{H}^{*\, \Sigma}
-\partial_{p}\mathcal{H}^{*\, \Lambda} \mathcal{H}^{\Sigma}]
=-\epsilon_{mnp}e^{-\mathcal{K}}\mathcal{Q}_{p}
[\mathcal{V}/X,(\mathcal{V}/X)^{*}]\, ,  
\end{equation}

\noindent
where $\mathcal{Q}_{p}[\mathcal{V}/X,(\mathcal{V}/X)^{*}]$ stands for
the pull-back of the K\"ahler 1-form substituting the $X^{\Lambda}$s
by the harmonic functions $\mathcal{H}^{\Lambda}$. For the $\bar{n}=2$
case, which can be embedded in pure $N=4,d=4$ supergravity, these
expressions were first found in Ref.~\cite{Bergshoeff:1996gg}.  We
stress that these functions are completely arbitrary and that there is
no further constraint on them. Different choices lead to solutions
describing different physical systems. In
Ref.~\cite{Bergshoeff:1996gg} the most general choice that leads to a
stationary, axisymmetric, asymptotically flat spacetime in the
$\bar{n}=2$ case was studied. These spacetimes correspond, in general,
to charged, rotating ``black holes'' (sometimes with singular
horizon), with NUT charge.

As one can see from Eq. (\ref{eq:PrepC}), the fact that we are dealing
with a quadratic prepotential implies that $\mathcal{C}_{ijk}=0$. This
then means that Eq. (\ref{eq:QueHorror}) is generically solved by
\begin{equation}
  \label{eq:QPrepPsi}
  \psi^{i} \; =\; e^{\mathcal{K}}\ \mathcal{G}^{ij^{*}}\ 
P_{j^{*}}(u,z^{*}) \; ,
\end{equation}
so that in the case of a quadratic prepotential there are only two
differential equations that need to be solved: eqs.
(\ref{eq:DetOmegaQ}) and (\ref{eq:EEuu}).


\subsection{$STU$-like models}
\label{sec:STUlike}

By an $STU$-like model, we mean a theory with a prepotential of the type
\begin{equation}
  \label{eq:STUPrep}
  \mathcal{F} \; =\; -d_{ijk}\frac{\mathcal{X}^{i}\mathcal{X}^{j}\mathcal{X}^{k}}{\mathcal{X}^{0}} \hspace{1cm} (i,j,k=1,\ldots ,n) \; ,
\end{equation}
where $d_{abc}$ is a totally symmetric tensor. The proper $STU$-model
is defined by $n=3$ and $d_{123}=1$ as the only non-vanishing
coefficients of $d$. The coefficients $d_{ijk}$ are related to the
$\mathcal{C}_{ijk}$ by
\begin{equation}
  \mathcal{C}_{ijk} \ =\ e^{\mathcal{K}}\ d_{ijk} \;\;\; ;\;\;\;
   e^{-\mathcal{K}} \ =\ 8\ d_{ijk}\ \Im m(Z^{i})\Im m(Z^{j}) \Im m(Z^{k}) \; .
\end{equation}
\par
In \cite{Shmakova:1996nz} Shmakova found a generic, conditional
solution to the stabilization equation for $STU$-like models. Writing
$\mathcal{I}^{T} = (p^{\Lambda}, q_{\Lambda})$, the stabilization
equations read $p^{\Lambda} = \mathrm{Im}(\mathcal{X}^{\Lambda})$ and
\begin{eqnarray}
  \label{eq:STUStab0}
  q_{0} & =& |\mathcal{X}^{0}|^{-4}\ d_{ijk}\ \mathrm{Im}\left( 
                         (\overline{\mathcal{X}}^{0})^{2}\mathcal{X}^{i}\mathcal{X}^{j}\mathcal{X}^{k}
             \right) \; , \\
  \label{eq:STUStaba}
  q_{i} & =& |\mathcal{X}^{0}|^{-2}\ d_{ijk}\ \mathrm{Im}\left(
                   \overline{\mathcal{X}}^{0}\mathcal{X}^{j}\mathcal{X}^{k}
             \right) \, .
\end{eqnarray}
Clearly the stabilization equations for $p^{\Lambda}$ are solved by the Ansatz
\begin{equation}
  \label{eq:STUStabsolvp}
  \mathcal{X}^{0} \ =\ p^{0}\left( i\ -\ Y_{0}\right) \;\;\; ,\;\;\;
  \mathcal{X}^{i} \ =\ Y^{i}\ -\ p^{i}Y_{0}\ +\ ip^{i} \; ,
\end{equation}
and plugging this Ansatz into the stabilization equation
(\ref{eq:STUStaba}), we can see that it can be solved by redefining
$Y^{i} = \sqrt{1+Y_{0}^{2}}\ \tilde{Y}^{i}$ iff a solution to
\begin{equation}
  \label{eq:STUStabConstraint}
  d_{ijk}\tilde{Y}^{j}\tilde{Y}^{k} \; =\; \textstyle{\frac{1}{3}}\ p^{0}q_{i} \; +\; d_{ijk}\ p^{j}p^{k} \; ,
\end{equation}
can be found. Assuming that such a solution exists, we can then
analyze Eq. (\ref{eq:STUStab0}) by direct substitution, as to find
\begin{equation}
  \label{eq:STUStabY0}
  1\ +\ Y_{0}^{2} \; =\; \frac{4\tilde{\Delta}^{2}}{
                             4\tilde{\Delta}^{2} 
                             - \left[
                                p^{0}p^{\Lambda}q_{\Lambda} + 2\Delta
                          \right]^{2}} \; , 
\end{equation}
where we have defined $\Delta = d_{ijk}p^{i}p^{j}p^{k}$ and
$\tilde{\Delta} = d_{ijk} \tilde{Y}^{i}\tilde{Y}^{j}\tilde{Y}^{k}$.
Armed with this knowledge one can then, using Eqs.
(\ref{eq:VbarVvgl}), calculate
\begin{equation}
  \label{eq:STUVVbar}
  \frac{1}{|X|^{2}} \; =\; \frac{4\sqrt{4\tilde{\Delta}^{2} 
                             - \left[
                                p^{0}p^{\Lambda}q_{\Lambda} + 2\Delta
                          \right]^{2}}
                           }{
                             p^{0}
                           } \; .
\end{equation}
It should be clear that Eq. (\ref{eq:STUStabConstraint}) acts as the
keystone for this construction.
\par
There are 2 cases for which a solution to Eq.
(\ref{eq:STUStabConstraint}) is near trivial to find. The first one is
the $STU$-model, so that $n=3$ and $d_{123}=1$ is the only
non-vanishing coefficient of $d$ where one can see that the general
solution reads
\begin{equation}
  \tilde{Y}^{i} \; =\; \sqrt{\frac{d_{ijk}\Upsilon_{j}\Upsilon_{k}}{4\Upsilon_{i}}} \;\;\;\; ;\;\;\;\;
  \Upsilon_{i} \;\equiv\; \textstyle{\frac{1}{3}}p^{0}q_{i} - d_{ijk}p^{i}p^{j} \; .
\end{equation}
Of course, the other case is when Eq. (\ref{eq:STUStabConstraint})
reduces to a purely quadratic equation, which happens when
$d_{ijk}=\delta_{ij}\delta_{jk}$, where for simplicity we have chosen
possible constants to be unity.  The solution then reads
\begin{equation}
  \tilde{Y}^{i} \; =\; \sqrt{\Upsilon_{i}} \; .
\end{equation}
With this knowledge and a choice for the harmonic functions, needed to
calculate $\omega$ through Eq. (\ref{eq:oidi}), the solution is fully
specified in the timelike case.
\par
The null case is in general a far harder nut to crack, and as one
might have suspected, we have been unable to find a generic solution
to Eq. (\ref{eq:QueHorror}) for the $STU$-like models.  A particular,
but non-trivial, solution we were able to find is for the case $n=1$,
$d_{111}=1$ and reads
\begin{equation}
  \psi^{1} \; =\; \left( 
                     a\ \Im m(t)^{-3} \ +\ ib\ \Im m(t)
                  \right)\ \partial_{z^{*}} t^{*} \; ,
\end{equation}
where we have used the notation $t=Z^{1}$.


\end{document}